\documentstyle[preprint,aps,epsf]{revtex}
\tighten

\newcommand{\grads}{\mbox{$^{\circ}$\hspace{-0.8ex}}}
\newcommand{\bold}[1]{\mbox{\boldmath $#1$}} 

\newcommand{\ie}{{\em i.e.}}                 
\newcommand{\etc}{{\em etc.}}                
\newcommand{\GeV}{{\rm GeV}}                 
\newcommand{\MeV}{{\rm MeV}}                 
\newcommand{\fm}{{\rm fm}}                   
\newcommand{\rme}{{\rm e}}                   
\newcommand{\k}{{\bf k}}                     
\newcommand{\beq}{\begin{equation}}
\newcommand{\eeq}{\end{equation}}
\newcommand{\beqar}{\begin{eqnarray}}
\newcommand{\eeqar}{\end{eqnarray}}
\newcommand{\bfig}{\begin{figure}}
\newcommand{\efig}{\end{figure}}
\newcommand{\quart}{\mbox{${1\over4}$}}         
\newcommand{\third}{\mbox{${1\over3}$}}         
\newcommand{\half}{\mbox{${1\over2}$}}          
\newcommand{\fourthird}{\mbox{${4\over3}$}}     

\begin{document}


\preprint{LBNL-42018}

\title{
DCC dynamics with the SU(3) linear sigma model%
\thanks{This work was supported by the Director, Office of Energy Research,
Office of High Energy and Nuclear Physics,
Nuclear Physics Division of the U.S. Department of Energy
under Contract No.\ DE-AC03-76SF00098.}}

\author{J\"urgen Schaffner-Bielich}
\address{
RIKEN BNL Research Center, Brookhaven National Laboratory\\
Upton, New York 11973}

\author{J\o rgen Randrup}
\address{
Nuclear Science Division, Lawrence Berkeley National Laboratory\\
University of California, Berkeley, California 94720}

\date{\today}
\maketitle

\begin{abstract} 
The SU(3) extension of the linear sigma model
is employed to elucidate the effect of including strangeness
on the formation of disoriented chiral condensates.
By means of a Hartree factorization,
approximate dispersion relations for the
18 scalar and pseudoscalar meson species are derived
and their self-consistent solution makes it possible to trace out
the thermal path of the two order parameters
as well as delineate the region of instability
within which spontaneous pair creation becomes possible.
The results depend significantly on the employed sigma mass,
with the highest values yielding the largest regions of instability.
An approximate solution of the equations of motion for the order parameter
in scenarios emulating uniform scaling expansions
show that even with a rapid quench only the pionic modes grow
unstable.
Nevertheless, the rapid and oscillatory relaxation of the order parameters
leads to enhanced production of both pions and (to a lesser degree)
kaons.
\end{abstract}

\draft


\section{Introduction}
\label{introduction}


High-energy collisions of atomic nuclei
may produce extended regions of space
within which chiral symmetry is temporarily nearly restored.
The subsequent non-equilibrium relaxation towards the normal vacuum
may then produce {\em disoriented chiral condensates} (DCC),
a coherent oscillation of the pion field 
that is manifested through anomalous pion multiplicity distributions
\cite{Anselm88,Anselm91,Blaizot:PRD46,bj,Rajagopal:NPB399}
and associated enhancements of electromagnetic processes
\cite{HW:ee,Boy:photons,KKRW}
(for reviews of the topic, see Refs.~\cite{Rajagopal:QGP2,BK:96,Bjorken}).
In order to assess the prospects for such DCC phenomena to actually occur
and be detectable above the background,
it is necessary to perform extensive dynamical calculations.
Most dynamical studies have been carried out
within the framework of the SU(2) linear $\sigma$ model
which incorporates the $\sigma$ and $\pi$ degrees of freedom
by means of an O(4) field
\cite{Rajagopal:NPB404,GGP93,Huang:PRD49,GM94,Blaizot:PRD50,Kluger,%
Boyanovsky:PRD51,Asakawa,Csernai,MM,Randrup:PRL,%
Biro,CGreiner,Camelia,Abada}.

While the omission of the strangeness degree of freedom
is justified in many nuclear studies, 
since the mass of the strange quark is large
compared with the temperatures typically encountered,
it may not be appropriate in the DCC context
because of the high temperature required
for the approximate restoration of chiral symmetry.
We therefore find it of interest to extend the treatment
to the strange sector and assess the effect of the DCC physics.

The resulting extended theory has a richer structure,
as two order parameters are now involved,
the light quark condensate {\em and} the strange quark condensate,
and the group of mesonic excitations has expanded from SU(2) to SU(3).
Moreover,
whereas the usual SU(2) treatment considers only the four fields
representing $\sigma$ and $\pi$ excitations,
the SU(3) model includes the eighteen meson fields
of the scalar and pseudoscalar nonets.

The incorporation of strangeness causes significant changes,
as will be discussed for thermal equilibrium
as well as the coupled dynamics of the two order parameters.
Moreover,
we explore the possibility that any of the mesonic quasi-particle modes
become unstable during the evolution.
The study is carried out within the framework of
the linear $\sigma$ model.
Although its SU(3) version has been known for a long time,
to our knowledge
it has not yet been utilized in connection with 
disoriented chiral condensates.

The presentation is organized as follows.
In Sect.~\ref{sec:su3} we introduce the SU(3) linear $\sigma$ model
and fix the parameters to the meson masses in the mean-field approximation.
The regions of instability of the various meson species
are then delineated in the plane of the two order parameters
and at zero temperature (Sect.~\ref{sec:regions}).
Subsequently, in Sect.~\ref{sec:temperature},
we study how the properties develop with temperature.
After having elucidated the equilibrium features,
we then, in section \ref{sec:dynamics},
consider various idealized dynamical scenarios 
that emulate what may be experienced in the rapidly expanding collision zone
following a high-energy nuclear collision.
Finally,
we briefly summarize our results.


\section{SU(3) linear $\sigma$ model}
\label{sec:su3}

The linear $\sigma$ model was first introduced by Gell-Mann and Levy
\cite{Gell60}. In a subsequent work, Levy extended this model from SU(2) to
SU(3) \cite{Levy67} based on the success of the Eightfold Way \cite{Gell64}
(for a review of the early work done in
SU(3) chiral models, see Ref.~\cite{Gasi69}).
It turns out that the model successfully accounts for the masses of the
pseudoscalar and scalar nonet, especially for the $\eta-\eta'$ splitting
by including a term that breaks U(3)$\times$U(3) symmetry but conserves the
SU(3)$\times$SU(3) symmetry (see, for example,
Refs.~\cite{Levy67,Sche71a,Carr71a,Carr71b,Olsh71}). 
The data for pion-pion and pion-kaon scattering can also be described 
\cite{Sche71b,Chan74} and 
the tree-level predictions are in agreement with Weinberg's
predictions \cite{Weinberg}.
Nevertheless, the nonlinear realization was preferred over the
linear one during the subsequent decades.

The linear description of chiral symmetry by means of
the scalar mesons as explicit degrees of freedom is supported
by the recent confirmation of the $\sigma$ meson \cite{Torn96},
which has led to the reappearance of the $\sigma$ meson
in the compilation by the Particle Data Group \cite{PDG96},
albeit with a very uncertain mass in the range of $400-1200$ MeV. 
Interestingly, 
the recent fit to the scalar mesons by T\"ornqvist \cite{Torn95} 
can in fact be interpreted as a model unitarization
based upon U(3)$\times$U(3) chiral symmetry \cite{Torn97}.

The SU(3) linear $\sigma$ model was formulated in the original paper
by Levy with the baryon octet \cite{Levy67}. 
Vector mesons can be included also as massive Yang-Mills fields 
\cite{Gasi69,Kaym85}. 
The trace anomaly can be
effectively accounted for by a glueball field \cite{Sche80}. 
The glueball field might be relevant for the chiral dynamics as
pointed out most recently for the 
SU(2) linear $\sigma$ model with quarks \cite{Mish98}. 
We leave the study of the effects of the glueball field 
on the DCC formation in SU(3) for future study.
This model, including baryons, vector mesons, and a dilaton field, 
has recently been applied to describe baryon masses as well as
nuclear matter and hypernuclei \cite{Papa98}. 

In the following we discuss the model for pseudoscalar and scalar mesons only,
as we are interested in meson-dominated environments,
such as those expected to be produced
in the central rapidity region
by relativistic nuclear collisions at RHIC.
Due to their higher masses,
vector mesons will become important only at higher temperatures
and will therefore not be considered here. 

The model used has a strong connection to QCD.
The order of the chiral phase transition of QCD was discussed using an
effective Lagrangian for arbitrary number of flavors
by Pisarski and Wilczek \cite{Pisa84}
using the linear SU(N) $\sigma$ model
which is the most general renormalizable Lagrangian
consistent with the symmetries of QCD. 
The chiral critical point was examined in \cite{Gavin94a} using the same
effective chiral Lagrangian for 2+1 flavors.
We will employ the same effective Lagrangian in the present study.


\subsection{The SU(3) meson states}

For the SU(2) version of the model,
there are only four mesonic modes,
the $\sigma$ isosinglet and the $\bold{\pi}$ isotriplet,
whereas the expanded SU(3) chiral model contains 18 different mesons,
the  nine pseudoscalar and the nine scalar mesons of the respective nonets.
The building block of the chiral SU(3) symmetric Lagrangian
is then the corresponding matrix M,
\begin{equation}
{\rm M}\ =\ \Sigma + i \Pi\ =\ \frac{1}{\sqrt{2}} \sum_{a=0}^8 \lambda_a 
 \left( \sigma_a+i\pi_a\right)
\quad .
\end{equation}
Here $\sigma_a$ and $\pi_a$ denote the U(3) components
of the scalar and pseudoscalar nonets, respectively,
and $\lambda_a$ denotes the standard Gell-Mann matrices
(including $\lambda_0=\sqrt{2\over3}\bold{I}$).

The explicit representation in terms of the physical states is
\begin{equation}
\Pi\ =\ \left( \begin {array}{ccc} 
 \frac{1}{\sqrt{2}}\pi^0+\frac{1}{\sqrt{6}}\eta_8+\frac{1}{\sqrt{3}}\eta_0 &
 \pi^+\phantom{00} & K^+ \cr 
 \pi^- &
 -\frac{1}{\sqrt{2}}\pi^0+\frac{1}{\sqrt{6}}\eta_8+\frac{1}{\sqrt{3}}\eta_0 &
 K^0 \cr 
 K^- & \bar{K}^0\phantom{00}
& -\frac{2}{\sqrt{6}}\eta_8+\frac{1}{\sqrt{3}}\eta_0
 \end{array} \right)
\quad , 
\label{eq:pmatrix}
\end{equation}
where $\eta_0$ and $\eta_8$ are mixed states of the $\eta$ and $\eta'$ mesons
and we note in particular the expressions for the pions and kaons
in terms of the Gell-Mann matrices,
\begin{eqnarray}
\label{eq:pioncomp}
&~&\pmatrix{\pi^+ \cr \pi^-} 
= \frac{1}{\sqrt{2}} \left(\lambda_1\pi_1\pm i \lambda_2\pi_2\right)
\quad , \qquad
\phantom{nn}\pi^0\phantom{ni} =\ \lambda_3\pi_3
\quad ,\\
\label{eq:kaoncomp}
&~&\pmatrix{K^+ \cr K^-} 
= \frac{1}{\sqrt{2}} \left(\lambda_4\pi_4\pm i \lambda_5\pi_5\right)
\quad , \qquad
\pmatrix{K^0 \cr \bar K^0} = \frac{1}{\sqrt{2}} \left(\lambda_6\pi_6\pm i
\lambda_7\pi_7\right) 
\quad .
\end{eqnarray} 
For the scalar nonet we have
\begin{equation}
\Sigma\ =\ \left( \begin{array}{ccc} 
 \frac{1}{\sqrt{2}} \left(\sigma+a_0^0\right) & a_0^+ & {K^*}^+ \cr
 a_0^- & \frac{1}{\sqrt{2}} \left(\sigma-a_0^0\right) & {K^*}^0 \cr 
 {K^*}^- & \bar{K}^{*0} & \zeta
 \end{array} \right) 
\quad ,
\label{eq:smatrix}
\end{equation}
where we have introduced the two order parameters, $\sigma$ and
$\zeta$, corresponding to the light quark and the strange quark condensate,
respectively. The $\zeta$ is in this notation a pure $s\bar{s}$ state,
{\em i.e.}\ ideal mixing is assumed. 
Nevertheless, as we will see, the $\sigma$ and $\zeta$ can still mix.


\subsection{SU(3)$\times$SU(3) chiral Lagrangian}

We start from the U(3)$\times$U(3) chirally symmetric
renormalizable Lagrangian for scalar and pseudoscalar mesons,
\begin{equation}
{\cal L}_{\rm U(3)}\ =\ 
\frac{1}{2} {\rm Tr} \left[\partial_\mu {\rm M} \partial^\mu
{\rm M}^\dagger \right] 
+ \frac{1}{2} \mu^2 {\rm Tr} \left[{\rm M} {\rm M}^\dagger \right]
- \lambda {\rm Tr} \left[{\rm M} {\rm M}^\dagger{\rm M} {\rm M}^\dagger \right]
- \lambda' \left({\rm Tr} \left[{\rm M} {\rm M}^\dagger\right]\right)^2
\quad .
\label{lag}
\end{equation}
Contrary to the SU(2) case,
there are now two coupling terms of fourth order in the fields,
with the strength parameters $\lambda$ and $\lambda'$.

In order to obtain a correct description of the meson masses,
it is necessary to include a term that breaks the U(3)$\times$U(3) symmetry
while still maintaining SU(3)$\times$SU(3) symmetry,
\begin{equation}\label{SU3}
{\cal L}_{\rm SU(3)}\ =\ 
c \det {\rm Tr} \left[{\rm M} + {\rm M}^\dagger \right] 
\quad .
\label{lagsu3}
\end{equation}
This term splits $\pi$ and $\eta'$ which are degenerate in U(3)$\times$U(3),
even in the presence of explicit symmetry breaking \cite{Carr71b}.
It is related to the U$_A$(1) anomaly of QCD and instanton-induced interactions
(see Ref.~\cite{Pisa84} and references therein). 
There are no additional independent terms that are invariant under
chiral SU(3)$\times$SU(3) transformations,
so any higher-order forms can be reexpressed
as functions of the three interaction terms already introduced.

The symmetry is explicitly broken
by an additional linear term in the Lagrangian,
\begin{equation}
\delta{\cal L}\ =\ f_0 \sigma_0 + f_8 \sigma_8\ =\
m_\pi^2 f_\pi \sigma
+ \left(\sqrt{2} m_K^2 f_K - \frac{1}{\sqrt{2}}m_\pi^2 f_\pi \right) \zeta\ ,
\label{lagesb}
\end{equation}
which makes the pseudoscalar mesons massive.
The second relation follows from the PCAC relations
for pions and kaons \cite{Papa98,Gasi69}.
The vacuum expectation values for the $\sigma$ and $\zeta$ fields are then
fixed by the pion and kaon decay constants,
\begin{eqnarray}
\sigma_{\rm vac} &=& f_\pi = 92.4 \mbox{ MeV }\ ,\\ 
\zeta_{\rm vac} &=& \sqrt{2} f_K - \frac{1}{\sqrt{2}}f_\pi 
= 94.5 \mbox{ MeV } \ ,
\end{eqnarray}
using the experimental values for the charged meson decay amplitudes, 
$f_{\pi^\pm} = 92.4 \pm 0.3$ MeV and $f_{K^\pm} = 113.0 \pm 1.3$ MeV.
We note that $\sigma_{\rm vac}$ and $\zeta_{\rm vac}$ are remarkably similar.


\subsection{SU(2) case}

In the limit of pure SU(2),
we consider only the $\sigma$ and $\pi$ fields in the
above expressions for the meson matrices (\ref{eq:pmatrix}-\ref{eq:smatrix}).
As the $\zeta$ field is assumed to be zero,
the SU(3) term (\ref{SU3}) vanishes.
Moreover,
the interaction terms proportional to $\lambda$ and $\lambda'$ are identical
because the relation
\begin{equation}
{\rm Tr} \left[{\rm M} {\rm M}^\dagger{\rm M} {\rm M}^\dagger \right] 
= \half \left({\rm Tr} \left[{\rm M} {\rm M}^\dagger\right]\right)^2
\end{equation}
holds for traceless matrices due to the Burgoyne's identity \cite{Cole85}.
The matrices are not traceless in the general U(3)$\times$U(3) 
case due to the presence of the singlet states. 
In SU(2) the Lagrangian then reduces to
\begin{equation}
{\cal L}_{\rm SU(3)\to SU(2)}\ 
=\ {\cal L}_{\rm kin} + \half \mu^2 \left(\sigma^2 + \pi^2\right) 
- \half (\lambda+2\lambda') \left(\sigma^2+\pi^2\right)^2 
+ f_\pi m_\pi^2\sigma \quad ,
\end{equation}
which is recognized as the SU(2) linear $\sigma$ model,
\begin{equation}
{\cal L}_{\rm SU(2)} = {\cal L}_{\rm kin} 
- \quart \lambda_{\rm SU(2)} \left(\sigma^2 + \pi^2 - v^2\right)^2 
+ f_\pi m_\pi^2\sigma\ ,
\label{eq:su2lag}
\end{equation}
if one assigns $\lambda_{\rm SU(2)}\equiv2\lambda+4\lambda'$
and $ v^2\equiv\mu^2/\lambda_{\rm SU(2)}$
and ignores the constant term $\quart\lambda_{\rm SU(2)}v^4$.
However,
since the SU(3) case has additional contributions
from the strange quark condensate,
the effective parameter values are shifted
and hence the values employed in the SU(2) linear $\sigma$ model
cannot be used as a guideline for SU(3).
Consequently,
the parameters of the SU(3)  model
will be fixed to the meson masses of the nonets. 


\subsection{Meson masses}

The remaining four parameters of the model, $\mu^2$, $\lambda$, $\lambda'$, and
$c$, are fixed by the meson masses in vacuum
which are given by the eigenvalues of the curvature tensor,
\begin{equation}
{\cal M}_{i,j}\ =\ -\frac{\delta^2{\cal L}({\bf\phi})}{\delta\phi_i\delta\phi_j}
\quad ,
\end{equation}
evaluated at the equilibrium point.
In the mean-field approximation,
the only mixings are between  $\sigma$ and $\zeta$
and between $\eta$ and $\eta'$.
The values of the model parameters are constrained by the fact that
the existence of a stable equilibrium
requires $\cal M$ to be positive definite,
{\em i.e.}\ the eigenvalues (the physical unmixed masses squared)
must be positive.
The expressions for the masses and the mixing angles 
are listed in the Appendix.

The easiest way is to fix the coupling
constants to the pion and kaon masses and the mass average
$\half(m_\eta^2+m_{\eta'}^2)$ \cite{Olsh71}. 
The $\eta$-$\eta'$ mass splitting is then predicted.
Then $\mu^2$ and $\lambda'$ are related
but can only be fixed by a choice for the scalar meson mass,
as discussed in detail in Ref.~\cite{Gavin94a}.
The parameter $\lambda'$
must be larger than $-2.85$ to ensure a positive value of $m_\sigma^2$. 
The mass parameter $\mu^2$ is only positive for $m_\sigma > 724$ MeV. 
As already noted,
$m_\sigma$ is not well known,
with current estimates ranging between 400 and 1200 MeV \cite{PDG96}.
Therefore,
we consider several cases within this range,
$m_\sigma= 400, 600, 800,$ and 1000 MeV. 
The parameters $\lambda$, $c$, and $\mu$
are analytically related to the pseudoscalar masses,
\begin{eqnarray}
\lambda\ &=&\
\frac{\left(m_\pi^2-\frac{1}{2}(m_\eta^2+m_{\eta'}^2)\right)
\left(2\zeta_{\rm vac}-\sqrt{2}\sigma_{\rm vac}\right) +
3\zeta_{\rm vac}\left(m_K^2-m_\pi^2\right) }
{\left(\sigma_{\rm vac}^2+4\zeta_{\rm vac}^2\right)\left(2\zeta_{\rm
vac}-\sqrt{2}\sigma_{\rm vac}\right)} 
\quad , \cr
c\ &=&\ \frac{m_K^2-m_\pi^2}
{\sqrt{2}\sigma_{\rm vac}-2\zeta_{\rm vac}} + 2 \lambda \zeta_{\rm
vac} \quad , \cr 
\mu^2\ &=&\ 4\lambda' \left(\sigma_{\rm vac}^2+\zeta_{\rm vac}^2\right)-m_\pi^2
+ 2\lambda\zeta_{\rm vac}^2 - 2 c \zeta_{\rm vac} \quad ,
\end{eqnarray}
and $\lambda'$ is then determined by $m_\sigma$.
The (unphysical) case $\sigma_{\rm vac}=\sqrt{2}\zeta_{\rm vac}$
yields degenerate masses for strange and non-strange mesons. 
These parameters are listed in Table \ref{tab:par}
together with the masses of the pseudoscalar and scalar meson nonets. 

The pseudoscalar mixing angle turns out to be $\theta_p=-5^\circ$.
For comparison, the Gell-Mann--Okubo mass formula gives $\theta_p=-10^\circ$
while experiment seems to favor $\theta_p=-20^\circ$ \cite{PDG96}. 
Nevertheless, the masses of $\eta$ and $\eta'$ are reproduced at the level
of $1$--$2\%$.
The parameters can be changed to get a pseudoscalar mixing angle of
$-10^{\circ}\hspace{-0.8ex}.4$
for a higher kaon mass and a lower kaon decay constant 
\cite{Gavin94a}. A more interesting approach to get a correct mixing angle has
been recently pointed out in \cite{Bura98} where the mass relations were
modified by the inclusion of the pseudoscalar decay constants.

The anomaly term (eq.\ (\ref{lagsu3})) is important for obtaining the correct
meson mass spectra,
especially for giving the $\eta'$ its large mass
and splitting the $\pi$ and $\eta$ masses.
In addition, the term mixes the $\eta$ and $\eta'$ meson states
away from the ideal mixing states $\eta_{ns}$ and $\eta_s$,
the former being a pure light-quark system
and the latter a pure $\bar ss$ state. 
There are arguments that the $SU(3)$ chiral phase transition
and the restoration of the $U_A(1)$ chiral symmetry 
does not necessarily coincide (scenario 1 considered by Shuryak \cite{Shuryak94}).
Then, a nonvanishing anomaly term would be responsible for giving the 
($\pi$, $\sigma$) and ($\eta_{ns}$, $a_0$) meson pairs different masses,
even in the $SU(3)$ chiral symmetric phase.
Their mass splitting is proportional to the $\zeta$ field,
hence is related to the finite strange quark mass.
Also the $\eta_{ns}$ gets a contribution from the anomaly term
due to the finite strange quark mass, even in the $SU(3)$ chiral limit,
but the $\eta_s$ does not. 
This could make the $\eta_s$ lighter than the $\eta_{ns}$,
as pointed out by Sch\"afer \cite{Schaefer96}.
On the other hand,
the $K$ and $K^*$ mesons are degenerate for a vanishing $\sigma$ field,
so their mass splitting from the anomaly term
is related only to the light quark masses. 

We recall that the situation for the scalar nonet is still
somewhat controversial (see Ref.~\cite[p.~355]{PDG96}
for a discussion of scalar mesons and their assignment to the scalar nonet). 
Weinstein and Isgur suggested, that the $f_0(980)$ as well as the
$a_0(980)$ are bound $K\bar{K}$ molecules \cite{Wein90}. The corresponding
meson states for the scalar nonet are higher in mass and are identified with
$f_0(1370)$ and $a_0(1450)$.
In this picture,
the mass of the $\zeta$ can be chosen to be close to $f_0(1370)$
but then the $a_0$ is too light (see Table \ref{tab:par}).
On the other hand, 
T\"ornqvist argues that all the lowest-lying scalar mesons 
are describable in a quark model and can be assigned to the scalar nonet;
the more massive scalar states are then excited states. 
In Table \ref{tab:par} we choose the latter picture for comparison. 
The mass of $a_0(980)$ is close to its experimental value.
The $\zeta$ is much heavier than the
$f_0(980)$, especially for the case $m_\sigma=1$ GeV. 
The mass of the $K^*$ seems to be far off compared to the $K^*(1430)$. 
Nevertheless, it was recently argued that there is evidence for a light scalar
$\kappa(900)$ \cite{Black98}.


\subsection{Idealized DCC's}

In the SU(2) case,
isospin invariance implies that the pion field vector
$(\pi_1, \pi_2, \pi_3)$
is equally likely to point in any particular direction,
$\hat{\bold{\pi}}=
(\cos\varphi_\pi\sin\vartheta_\pi,
\sin\varphi_\pi\sin\vartheta_\pi,\cos\vartheta_\pi)$.
If the iso-orientation of the pion field is the same
throughout the entire source,
then all the radiated pions emerge with the same isospin alignment
in each separate event.
It then follows that the resulting {\em neutral pion fraction}
$R_\pi\equiv N_{\pi^0}/N_\pi = \cos^2\vartheta_\pi$ has the distribution
$P_\pi(R_\pi)=1/2\sqrt{R_\pi}$
\cite{Anselm91,Blaizot:PRD46,bj,Rajagopal:NPB399}.
This distribution differs significantly from
the narrow peak around one third that would be expected
if each individual pion had a random isospin orientation
and it has therefore been proposed as a suitable DCC signal.

This situation remains the same in the SU(3) case
with regard to the three-dimensional pion subspace.
But, in addition,
there is also invariance in the four-dimensional subspace
of the kaon field ($K^+,K^-,K^0,\bar{K}^0$),
{\em i.e.}\ the direction of the kaon field vector
$(\pi_4, \pi_5, \pi_6, \pi_7)$,
\begin{equation}
\hat{\bold{K}}=
(\cos\varphi_K\sin\vartheta_K\sin\psi_K,
\sin\varphi_K\sin\vartheta_K\sin\psi_K,
\cos\vartheta_K\sin\psi_K,\cos\psi_K)\ ,
\end{equation}
is equally likely to point in any direction.
Thus, by analogy with the pions,
it is natural to consider the {\em neutral kaon fraction},
\begin{equation}
R_K\    \equiv\ {N_{K^0} + N_{\bar{K}^0} 
\over   N_{K^+} +N_{K^-} + N_{K^0} + N_{\bar{K}^0} }\
=\ 1-\sin^2\vartheta_K \sin^2\psi_K\ . 
\end{equation}
Under the similar assumption of full alignment of the kaon field
in each distinct event,
it is elementary to calculate the corresponding distribution,
\begin{eqnarray}
P_K(R_K) &=&  {1\over2\pi^2} \int_0^\pi d\psi_K \sin^2\psi_K
\int_0^\pi d\vartheta_K \sin\vartheta_K
\int_0^{2\pi} d\varphi_K\
\delta(R_K-1+\sin^2\vartheta_K \sin^2\psi_K)\\
&=& {1\over\pi} \int_0^\pi d\psi_K\sin\psi_K\
\theta(R_K-\cos^2\psi_K)\
[R_K-\cos^2\psi_K]^{-{1\over2}}\\
&=& {2\over\pi} \int_0^{\sqrt{R_K}} d\cos\psi_K\
[R_K-\cos^2\psi_K]^{-{1\over2}}\ =\ 1\ .
\end{eqnarray}
Hence, all values of the neutral kaon fraction $R_K$ are equally likely,
whereas a statistical emission would yield a narrow distribution
peaked at one half.
As was the case for $P_\pi(R_\pi)$,
the result for $P_K(R_K)$
is independent of any explicit chiral symmetry breaking,
as it relies only on the invariance with regard to transformations
among the four kaon states.

Contrary to the neutral pions,
neutral kaons may decay by weak interactions into hadrons,
$K^0_S\to\pi^+\pi^-$,
which can be detected in heavy-ion experiments by standard techniques.
Although event-by-event reconstruction is impractical,
fluctuation analysis of the abundance of neutral and charged kaons 
has been proven feasible with the present heavy-ion experiments at the SPS.
An enhancement of the ratio near zero or one,
or at least a broadening of $P_K(R_K)$,
might be visible and would suggest the occurrence of non-equilibrium effects 
of the DCC type.


\section{Regions of stability at zero temperature}
\label{sec:regions}


In this section we examine the regions of instability
for the pseudoscalar mesons
in the plane of the two order parameters $\sigma$ and $\zeta$,
at zero temperature.
This information is particularly relevant in the idealized quench scenario,
in which the initially hot system,
in which chiral symmetry is at least partially restored,
is suddenly cooled down.
Generally,
the effective meson masses depend on both condensates
and it is thus more complicated to map out the occurrence of instabilities.

Lattice gauge calculations indicate that
although the light-quark condensate does not entirely vanish
in the chirally restored phase it is considerably reduced. 
The strange-quark condensate follows this trend
but its reduction is less pronounced
due to the higher mass of the strange quark \cite{Kogut91}.
The different behavior of the strange-quark condensate is ensured
in the linear SU(3) $\sigma$ model with the inclusion of the $\zeta$ field.
The mean-field expressions for the effective meson masses
as functions of the order parameters and the thermal field fluctuations
are given in the Appendix. 
In the following, we study only the pseudoscalar mesons
as they are directly measurable in heavy-ion experiments.

In the mean-field approximation \cite{Randrup:PRD},
the quasi-particles satisfy the Klein-Gordon dispersion relation,
$\omega_k^2=k^2+m^2$.
Thus, whenever a negative value of $m^2$ occurs,
the frequency $\omega_k$ is imaginary
for sufficiently small values of the wave number $k$,
leading to an exponential growth of the corresponding field component.
This phenomenon is familiar from the discussion of supercritical fields
and it manifests itself as an enhancement of the number of soft mesons
of the affected kind.
Therefore it is of interest to delineate the regions of stability
for the various meson species.

\begin{figure}[htbp]
\begin{center}
\leavevmode
\epsfxsize=0.7\textheight
\epsfbox{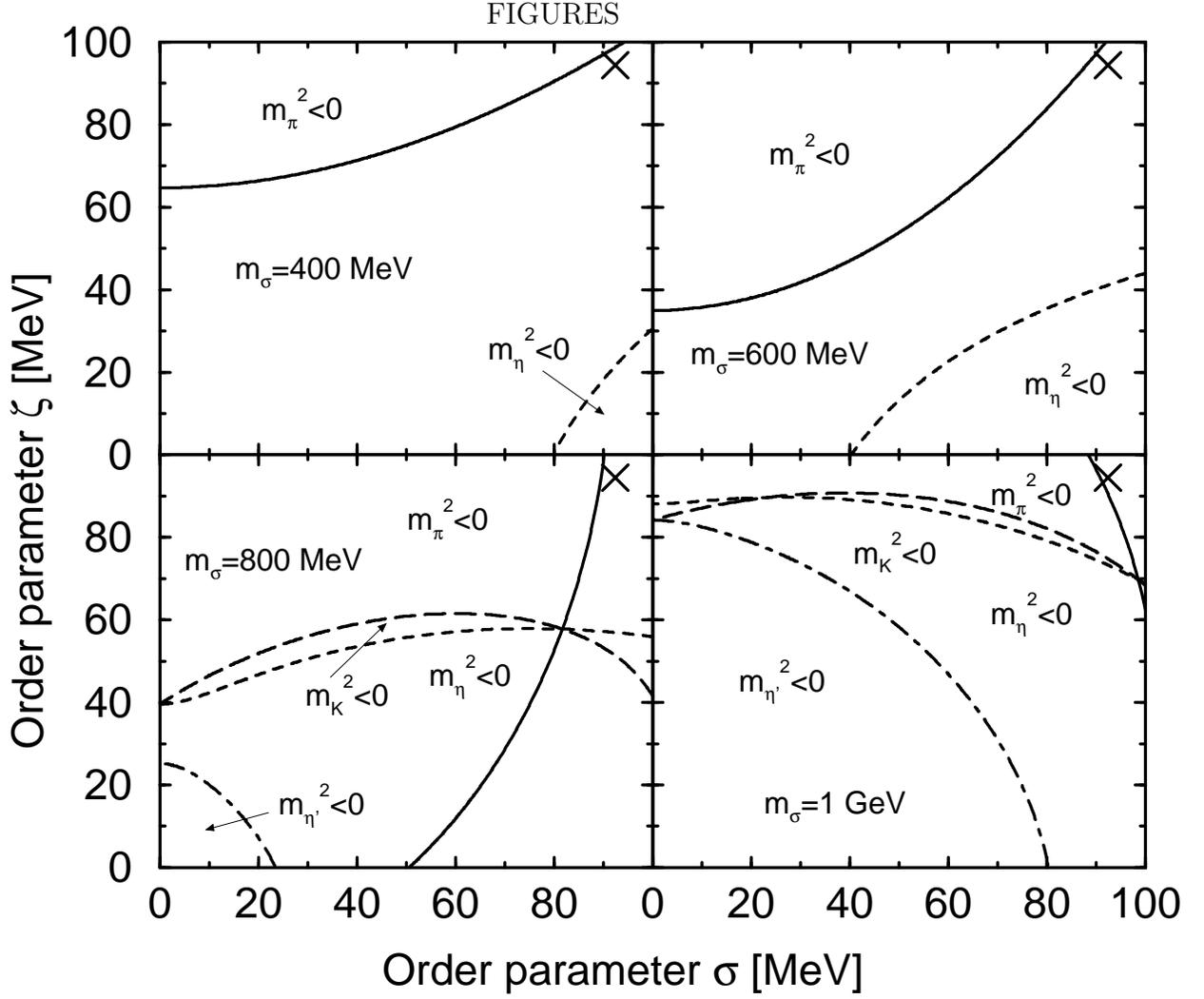}
\end{center}
\caption{The critical boundaries for the pseudoscalar mesons
in the plane of the order parameters
$\sigma=\langle\bar{q}q\rangle$ and $\zeta=\langle\bar{s}s\rangle$.
The cross indicates the vacuum equilibrium point
$(\sigma_{\rm vac},\zeta_{\rm vac})$
where all meson masses are positive.
The results were obtained with four different values of $m_\sigma$
as indicated.
The curves delineate the various critical boundaries,
where the respective effective mass vanishes.}
\label{Fig1}
\end{figure}

Figure \ref{Fig1} shows the critical boundaries
for the pseudoscalar mesons for the four selected values of $m_\sigma$.
We see that $m_\pi^2$ is negative in the upper-left part
while $m_\eta^2$ is negative in the lower-right part.
For the relatively small mass of 400~MeV
the $\pi$ and $\eta$ are the only mesons exhibiting instabilities
and the corresponding supercritical regions are fairly small.
In particular,
there is no combination of the two order parameters $\sigma$ and $\zeta$
that would cause the kaons to become unstable.
These general features remain true also at $m_\sigma=600~\MeV$,
although the regions of $\pi$ and $\eta$ instability
have expanded considerably.

This expansion continues steadily as $m_\sigma$ is increased further
and by 800~MeV the situation has changed dramatically,
as the model parameter $\mu^2$ has turned positive.
Instabilities are now possible for all four pseudoscalar mesons
and occur simultaneously in a small region near the origin
where full chiral symmetry is approximately restored. 
However,
it should be noted that the instabilities of $K$, $\eta$, or $\eta'$
occur only for considerably reduced values of the strange order parameter
$\zeta$.
Thus, a reduction of only the light-quark condensate $\sigma$ 
could produce only the standard pionic instability.
Moreover,
the critical boundaries for the $\eta$ and $K$ mesons
are quite close to one another.

Finally,
for $m_\sigma=1$ GeV,
an imaginary pion mass is reached already after only a slight reduction
of either $\sigma$ or $\zeta$.
Furthermore,
the kaon is now unstable over a wide order-parameter region
and the critical boundary for $\eta$ is again close to the kaon boundary.
We also note that the critical boundary for $\eta'$ has expanded considerably
so that this meson may exhibit instability as well
even if chiral symmetry has been only partly restored.


\section{Finite temperature}
\label{sec:temperature}


We now study how the properties of the system develops
as the temperature is raised.
Statistical equilibrium of the SU(3) linear $\sigma$ model
can be treated in the mean-field approximation
along the lines developed for the SU(2) case \cite{Randrup:PRD}.
The local field strength is then decomposed
into the (smooth) order parameter and the residual fluctuations,
\begin{equation}
\sigma \longrightarrow \langle\sigma\rangle + \delta\sigma \quad \mbox{and}
\quad  
\zeta \longrightarrow \langle\zeta\rangle + \delta\zeta \quad.
\end{equation}
The insertion of this decomposition into the general equation of motion
then leads to separate equations of motion for the order parameters
and the individual quasi-particle fields.
The trilinear fluctuation terms are included
by the Hartree factorization technique, 
\begin{equation}
\delta\phi_i \delta\phi_j \delta\phi_k\ \longrightarrow\
\langle\delta\phi_i \delta\phi_j \rangle\delta\phi_k \;+\;
\langle\delta\phi_j \delta\phi_k \rangle\delta\phi_i \;+\;
\langle\delta\phi_k \delta\phi_i \rangle\delta\phi_j
\end{equation}
where $\langle\cdot\rangle$ denotes the thermal average,
\begin{equation}
\langle\delta\phi_i \delta\phi_j \rangle\ =\
\delta_{ij}\ {1\over\Omega} {\sum_\k}'
 \frac{1}{\omega_k} \frac{1}{\exp\left(\omega_k/T\right)-1}\ 
=\ \delta_{ij} \frac{m_i T}{2\pi^2} \sum_{n=1}^\infty \frac{1}{n} K_1 
 \left(\frac{n m_i}{T}\right)\ ,
\label{eq:fluct}
\end{equation}
in the diagonal representation.
The sum is over all finite wave numbers $\k$
and the associated energy is $\omega_k=\sqrt{m_i^2+k^2}$
where $m_i$ is the appropriate temperature-dependent effective mass
for the meson species $i$,
as obtained by solving the coupled equations selfconsistently.
We need to take into account only the thermal fluctuations of
$\pi$, $K$, $\sigma$, and $\eta$ mesons,
as they are the lightest and thus have the largest field fluctuations.
The other mesons ($\eta'$, $a_0$, $K^*$, $\zeta$)
have masses near or above one GeV
and the next level of refinement would then rather involve
the fluctuations of the vector meson fields. 

The thermal equilibrium values
of the two order parameters are then determined by the resulting
 equations of motion for $\langle\sigma\rangle$ and $\langle\zeta\rangle$, 
\begin{eqnarray}\label{eq:eomsigma} 
\mu_\sigma^2 \langle\sigma\rangle &=& 
f_\pi m_\pi^2 + c \left(\frac{4}{3\sqrt{2}} \langle\delta\eta^2\rangle
+ \frac{1}{\sqrt{2}} \langle\delta K^2\rangle
 + 2 \langle\delta\sigma\delta\zeta\rangle\right)\\ \nonumber
&~&+ \left(\sqrt{2} \lambda' \langle\delta K^2\rangle
- 8 \lambda \langle\delta\sigma\delta\zeta\rangle \right)\langle\zeta\rangle\ ,
\\ \label{eq:eomzeta}
\mu_\zeta^2 \langle\zeta\rangle &=& 
\sqrt{2} f_K m_K^2 - \frac{1}{\sqrt{2}} f_\pi m_\pi^2 
+ c \left(\langle\delta\pi^2\rangle+\langle\sigma\rangle^2
+\langle\delta\sigma^2\rangle
-\frac{1}{3}\langle\delta\eta^2\rangle \right)\\ \nonumber
&~& + \left(\sqrt{2} \lambda' \langle\delta K^2\rangle 
- 8 \lambda \langle\delta\sigma\delta\zeta\rangle \right)\langle\sigma\rangle
\quad ,
\end{eqnarray}
where the effective masses for the order parameters are given by
\begin{eqnarray} \label{eq:musigma} 
\mu_\sigma^2 &=& 
-\mu^2 - 2c\langle\zeta\rangle + 2\lambda
\left( \langle\delta\pi^2\rangle + \langle\sigma\rangle^2 
+ 3\langle\delta\sigma^2\rangle 
+ \frac{1}{3} \langle\delta\eta^2\rangle + \langle\delta K^2\rangle \right)\ ,
\\ \nonumber
&~&+ 4\lambda' 
\left( \langle\delta\pi^2\rangle + \langle\sigma^2\rangle 
+ 3\langle\delta\sigma^2\rangle + \langle\zeta^2\rangle +
\langle\delta\zeta^2\rangle + \langle\delta\eta^2\rangle
+ \langle\delta K^2\rangle\right) \\  \label{eq:muzeta}
\mu_\zeta^2 &=& 
-\mu^2+4\lambda \left(\langle\zeta^2\rangle + 3\langle\delta\zeta^2\rangle 
+ \frac{2}{3} \langle\delta\eta^2\rangle 
+ \langle\delta K^2\rangle\right)\\ \nonumber
&~&+4\lambda' 
\left(\langle\delta\pi^2\rangle + \langle\sigma\rangle^2 
+ \langle\delta\sigma^2\rangle + \langle\zeta\rangle^2 +
3\langle\delta\zeta^2\rangle + \langle\delta\eta^2\rangle
+ \langle\delta K^2\rangle\right)
\quad .
\end{eqnarray}
The mixing of the field fluctuations $\delta\sigma$ and $\delta\zeta$ 
can be treated to a sufficient degree of approximation by the replacements 
\begin{eqnarray}\nonumber
\langle\delta\sigma^2\rangle\ &\longrightarrow&\  
\langle\delta\sigma^2\rangle \cos^2{\theta_s}\ ,\\ \label{eq:fluctmix}
\langle\delta\zeta^2\rangle\ &\longrightarrow&\
\langle\delta\sigma^2\rangle \sin^2{\theta_s}  \ ,\\ \nonumber
\langle\delta\sigma\delta\zeta\rangle\ &\longrightarrow&\
\langle\delta\sigma^2\rangle \sin{\theta_s} \cos{\theta_s}\ ,
\end{eqnarray}
where we have neglected fluctuations of the heavier $\zeta$ meson.
Mixing between $\eta$ and $\eta'$ is ignored for the thermal fluctuations
as the mixing angle turns out to be near zero even at high temperatures. 
The final expressions for the meson masses can be found in the Appendix.

In order to simplify the notation,
we omit the brackets $\langle\cdot\rangle$ in the following
so that the order parameters will be denoted by $\sigma$ and $\zeta$
and the thermal field fluctuations by $\delta K^2$ and so on.
The coupled equations for the order parameter
(\ref{eq:eomsigma},\ref{eq:eomzeta}), 
the meson masses (see Appendix), and the
fluctuations (\ref{eq:fluct},\ref{eq:fluctmix}) 
are solved iteratively with an overdamped Newton method.

We note in passing, that we assume that the coupling constants of the
effective interaction is temperature independent. This is a strong assumption,
especially for the anomaly term where one knows that the corresponding
coupling constant should be much smaller at the chiral phase transition
\cite{Shuryak94}. Effects should be especially seen for the $\eta/\eta'$ mesons
\cite{Kapusta96,Huang96} and could even lead to parity-violating effects as
pointed out recently \cite{Kharzeev98}.
How the coupling constants scale with temperature is not
precisely known. There exists estimates but they can be applied in
principle only for a quark-gluon plasma. Therefore, we leave this interesting 
question for future investigations.


\subsection{Temperature dependence of the order parameters}

\begin{figure}[htbp]
\begin{center}
\leavevmode
\epsfxsize=0.7\textheight
\epsfbox{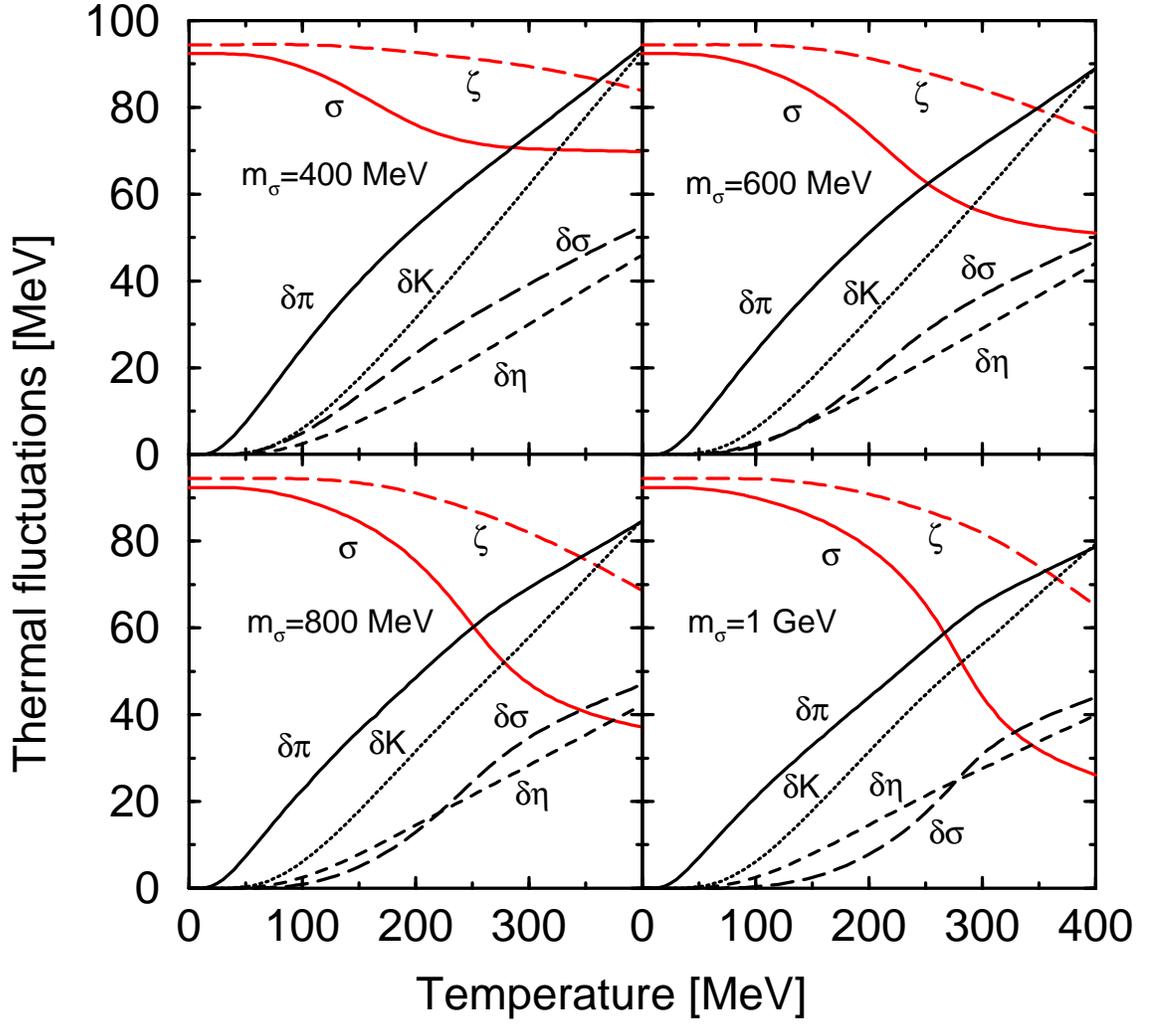}
\end{center}
\caption{The temperature dependence of the $\pi$, $K$, $\eta$, and $\sigma$
field fluctuations in thermal equilibrium,
together with the equilibrium values of the two order parameters
$\sigma$ and $\zeta$,
for $m_\sigma$ = 400, 600, 800, and 1000 MeV.}
\label{Fig2}
\end{figure}

Figure \ref{Fig2} shows the temperature dependence of the total
fluctuations of the $\pi$, $K$, $\sigma$, and $\eta$ quasi-particle fields
for the various values of $m_\sigma$
(\ie\ $\delta\pi^2\equiv\langle\delta\pi_1^2\rangle
+\langle\delta\pi_2^2\rangle+\langle\delta\pi_3^2\rangle$, \etc).
As expected,
the pion fluctuations are dominant at low temperatures due to their small mass,
and the other contributions can be safely neglected for $T<50$ MeV.
The situation changes drastically at higher temperatures
as the strange degrees of freedom are activated.
The kaon fluctuations constitute a 20\% correction at $T=100$ MeV
and are already half the pion contribution at $T=150$ MeV
and can thus no longer be neglected.
The kaon fluctuations become remarkably similar to the ones of the pions
at higher temperatures and contribute equally at $T=400$ MeV. 
Note that the degeneracy factor for the kaon is four thirds times
that of the pion so that the kaon contribution would dominate
if the masses of the pion and the kaon were equal.
The thermal fluctuations do not change significantly with $m_\sigma$,
except of course for those of the $\sigma$ field itself
which for large $m_\sigma$ grow rapidly in the $\sigma$ crossover region
as the effective $\sigma$ mass experiences a pronounced minimum
(see Fig.~\ref{fig:masses}).

The figure also displays the thermal evolution of the two order parameters.
They exhibit a smooth crossover
from their large (and nearly equal) vacuum values
towards ever smaller values at high temperatures.
The fastest change for $\sigma$ occurs around $T=210\ \MeV$.
The decrease of the strange order parameter $\zeta$ is much less pronounced,
as a consequence of the larger kaon mass,
and the corresponding crossover does not occur until above 400 MeV.

It is evident that the degree of chiral symmetry restoration achieved
depends sensitively on the adopted value of $m_\sigma$.
For the smaller values the order parameters decrease only moderately.
For example, at $T=400~\MeV$ $\sigma$ is still above 50~MeV,
for the commonly employed value of $m_\sigma=600~\MeV$
whereas it has dropped below 30~MeV for $m_\sigma=1~\GeV$.
The more complete symmetry restoration at higher $m_\sigma$ values
is due to the fact that the parameter $\mu^2$ has now become positive and the
coupling constant $\lambda'$ has increased.
The former feature affects the spontaneous symmetry breaking
at zero temperature,
while the latter enhances the temperature dependence of the meson masses. 
On the basis of these observations,
we expect that the most dramatic effects of the extension to SU(3)
will occur for large values of $m_\sigma$. 

\begin{figure}[htbp]
\begin{center}
\leavevmode
\epsfxsize=0.7\textheight
\epsfbox{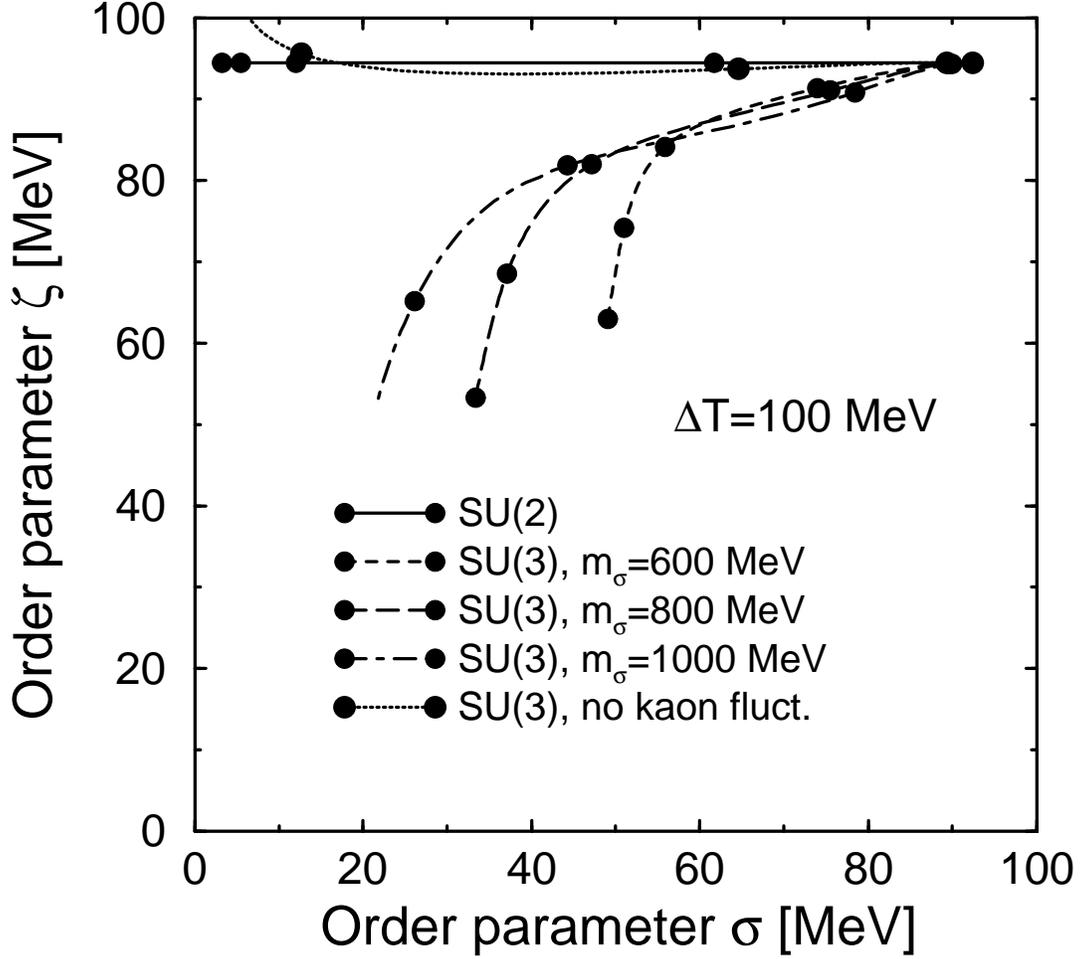}
\end{center}
\caption{The thermal path of the two order parameters
for various cases of interest, as indicated.
The solid dots are plotted in temperature steps of 100 MeV.
All paths start from the vacuum point at the upper right corner
and the points for $T=100~\MeV$ are still very close to the vacuum point.} 
\label{fig:orderpar}
\end{figure}

\begin{figure}[htbp]
\begin{center}
\leavevmode
\epsfxsize=0.7\textheight
\epsfbox{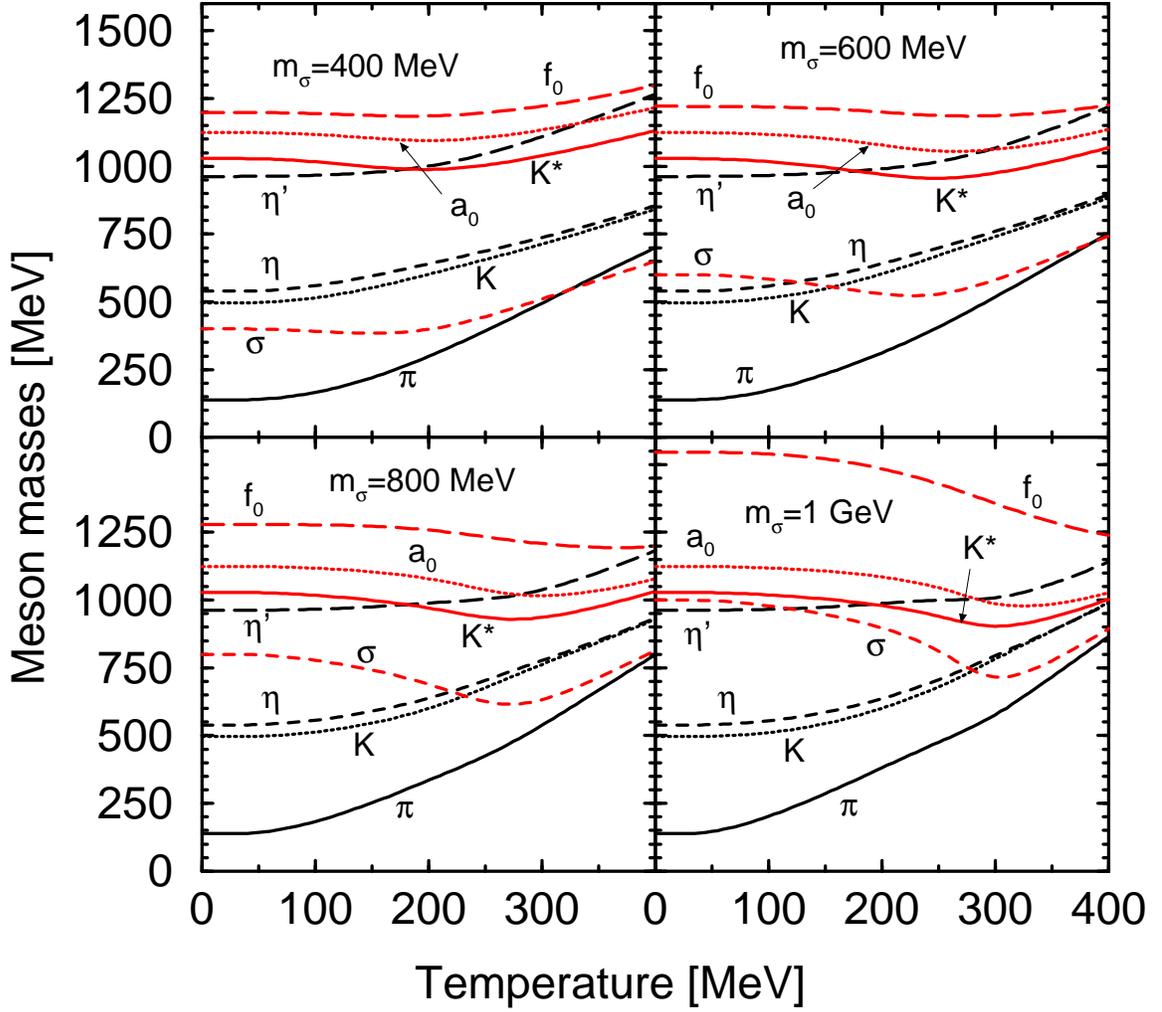}
\end{center}
\caption{The temperature dependence of the effective masses
of the pseudoscalar ($\pi$, $K$, $\eta$, $\eta'$)
and scalar ($\sigma$, $\zeta$, $K^*$, $a_0$) meson nonets
for the four values of $m_\sigma$ studied, as indicated.} 
\label{fig:masses}
\end{figure}

The temperature dependence of the two order parameters
is well illustrated by the thermal path of the equilibrium point
in the $(\sigma,\zeta)$ plane,
as shown in Fig.~\ref{fig:orderpar} for $m_\sigma=600,800,1000~\MeV$.
The solid circles mark the progression of the path
in temperature intervals of 100 MeV.
In addition,
we also show the result of omitting the kaon fluctuations
as well as the result for the pure SU(2) case.
In the latter case,
there is obviously no change in the strange order parameter
and we note that $\sigma$ reaches very small values
for temperatures above $T=300$
(the last two dots are for temperatures of 400 and 500 MeV).
The SU(3) case without kaon fluctuations is close to the pure SU(2) case
but differs from the full SU(3) calculation.
In particular, $\zeta$ hardly changes in the absence of kaon fluctuations. 
The paths for the full SU(3) calculation
develop very slowly at low temperatures
(note that the points for $T=100$ MeV are quite close to the vacuum point),
which reflects the above finding that
the order parameters change little over that temperature interval.
In the intermediate temperature range of $T=150-300$ MeV
both order parameters decrease approximately linearly
and the path follows a nearly straight line.
At higher temperatures, $\sigma$ quickly saturates
while $\zeta$ is only now beginning to drop faster.
At these high temperatures,
the kaon fluctuations now provide the dominant contribution to the meson masses.
We see that the inclusion of kaon fluctuations
({\em i.e.}\ going from SU(2) to SU(3))
has a dramatic effect on the thermal evolution of the order parameters. 

It is interesting to compare our findings for the full SU(3) case 
with related previous work.
The linear SU(3) $\sigma$ model was solved in the $1/N_f$ expansion
by means of the saddle-point approximation
for finite temperatures \cite{Meyer96}.
It was found that the light-quark condensate drops sharply at 180-190 MeV,
while the strange-quark condensate has changed by only 20\% at $T=300$ MeV
as compared to its vacuum value. 
On the other hand, 
we find that the changes of the order parameters
depend sensitively on the chosen parameter values,
especially the value for $m_\sigma$.
The heavier the value of $m_\sigma$ is,
the larger is the drop of both order parameters
as the temperature is increased and the higher is the critical region
in which the cross over occurs.
The change of the strange-quark condensate shown in Ref.\ \cite{Meyer96}
is smaller than in our mean-field results,
while it is the reverse situation for the light-quark condensate. 

There exists only one lattice calculation of the light and strange
quark condensates for $N_f=2+1$ \cite{Kogut91}.
The light-quark condensate drops more than 90\% through the transition region,
while the strange-quark condensate decreases only moderately (by 30\%)
over the same temperature region. 
Our calculations yield a less dramatic behavior,
especially for a low $\sigma$ vacuum mass,
which thus seems to favor a high vacuum value for $m_\sigma$.
For the case $m_\sigma=1$ GeV,
we obtain a change of only 70\% for $\sigma$
and about 30\% for $\zeta$.
Still, the $\sigma$ condensate changes less than indicated by lattice data,
while the strange condensate follows the trend seen on the lattice
but shifted towards higher temperatures.
However,
this comparison should be taken only as suggestive,
as we are using an effective model.


\subsection{Meson masses at finite temperature}

The effective meson masses will exhibit a temperature dependence
due to the thermal decrease of the order parameters
and the associated growth of the field fluctuations
(the explicit expressions for the meson masses,
including the contributions from thermal fluctuations,
are given in the Appendix).
The effective masses are obtained in connection with the above
self-consistent solution of the coupled equations carried out for the purpose
of determining the temperature dependence of the order parameters.

Figure \ref{fig:masses} shows the resulting masses
for the pseudoscalar nonet ($\pi$, $K$, $\eta$, $\eta'$)
and the scalar nonet
($\sigma$, $a_0$, $K^*$, $f_0$).
For temperatures below 80~MeV 
there is hardly any change in the meson masses at all.
For intermediate temperatures, $80<T<230$ MeV,
the $\pi$ mass as well as the $K$ and $\eta$ masses
start to increase approximately linearly.
The $\eta'$ mass is also increasing with temperature
but relatively less so than the other pseudoscalar meson masses.
By contrast,
all scalar meson masses decrease with temperature through this range,
since the decrease of the order parameters
bring them closer to their respective critical boundaries.

For $m_\sigma=600~\MeV$,
the $\sigma$ meson mass drops below the $K$ and $\eta$ masses
and reaches its lowest value of $520~\MeV$ at $T=230~\MeV$.
The $a_0$ and $K^*$ mesons masses change similarly,
losing about 70~MeV of mass, 
but the $f_0$ mass drops only 35~MeV from its free value. 
At even higher temperatures, $T>230~\MeV$,
the thermal fluctuations dominate all the meson masses
and they now increase linearly with temperature.
The slope for the pseudoscalar mesons is steeper
than that for the scalar mesons.
Especially the pion mass changes remarkably at high temperatures,
reaching 750~MeV at $T=400~\MeV$.
The $\sigma$ meson becomes degenerate with the pion at around that temperature.
The $K$ and $\eta$ masses are close to one another
and reach values just below 900~MeV at $T=400~\MeV$.
The $\eta'$ mass is rising so rapidly with temperature
that it becomes the heaviest meson of all
at temperatures above 400~MeV. 

For $m_\sigma=800$ MeV,
the pseudoscalar meson masses show a similar behavior,
except that the $\eta'$ is rising less steeply through the transition region.
All scalar meson masses are again dropping 
as the order parameters are dropping towards chiral symmetry restoration,
but it is now more pronounced.
The $\sigma$ mass is falling by nearly 25\%
to $m_\sigma=615~\MeV$ at $T=270~\MeV$.
The masses of $a_0$ and $K^*$ drop from their free values by $100~\MeV$.
Furthermore, the $f_0$ loses about 85~MeV,
but the minimum is reached at the higher temperature of $T=370~\MeV$.
Again, the $\pi$ and $\sigma$ masses,
as well as those of $\eta$ and $K$,
are all becoming very similar behavior above 300~MeV. 

The changes for the scalar meson masses are even more pronounced
for $m_\sigma=1~\GeV$.
The $\sigma$ mass now drops by 28\%
and reaches its minimum value of 716~MeV at $T=305~\MeV$. 
The $a_0$ and $K^*$ are dropping 125~MeV and 146~MeV, respectively.
The minimum of the $K^*$ mass is located at a slightly higher temperature,
$T=325~\MeV$, due to its strong coupling to the strange order parameter $\zeta$
which starts decreasing at a higher temperature than $\sigma$.
The $f_0$  now shows a significant mass change,
having decreased from 1550~MeV to 1240~MeV at $T=400~\MeV$.
If continuing to higher temperatures,
a minimum mass of 1220~MeV would be reached at $T=440~\MeV$.
Here the $f_0$ mass is strongly influenced by the strange order parameter,
hence the mass minimum is shifted to higher temperatures
as compared to the other scalar mesons.
The pseudoscalar meson masses are showing a steeper rise with temperature,
as compared to the other cases with a lower vacuum value of $m_\sigma$.
The $\eta'$ mass shows a different behavior,
as it remains rather unchanged up to $T=300~\MeV$,
before starting its growth.
At $T=400~\MeV$, the masses of $K$, $\eta$, $a_0$, and $K^*$
are closely clustered around 1~GeV.
The $\eta'$ and $f_0$ masses appear to converge at higher temperatures,
signaling the onset of full chiral symmetry restoration. 

Comparing the results obtained with the different $\sigma$ masses,
we observe some general trends.
First, the pions, as well as the other pseudoscalar mesons, 
grow more massive at high temperatures;
there are thus fewer medium-modified pions in a thermalized system
at high temperatures as compared to a free gas at the same $T$. 
Second, the effective $K$ and $\eta$ mesons
become approximately degenerate with the effective $\pi$ meson.
This will enhance the yield ratios to $K/\pi\approx\fourthird$ and
$\eta/\pi\approx\third$ as compared with the free-gas estimate of
$K/\pi\approx\fourthird \rme^{-\Delta m/T}\approx 0.5$ 
and $\eta/\pi\approx 0.1$
for $T=400$ MeV.

At high temperature,
the thermal fluctuations of the vector mesons can no longer be neglected.
Of course, their impact on the results presented depend
on the change of their masses with temperature.
If the vector meson masses stay near their vacuum values up to $T=300~\MeV$,
or perhaps drop with temperature,
the corresponding thermal fluctuations will grow important already
in the transition region for the order parameter $\sigma$.
On the other hand,
if the vector meson masses increase with temperature,
as does the $\eta'$,
then their contributions can possibly be neglected
over the temperature range shown,
as they are too heavy to become significantly agitated.

The thermal $a_0$ and $K^*$ fluctuations
become important at high temperatures, say $T>300$ MeV,
as their masses are then close to those of $\eta$ and $K$, respectively.
As the thermal masses are in the range of 1~GeV,
vector mesons should be also included and might constitute a bigger correction
term at very high temperatures.
Therefore,
we omitted the $a_0$ and $K^*$ fluctuations.
It might be interesting to study vector mesons within the linear $\sigma$ model
and they should clearly be included in studies of the high-temperature behavior
of pseudoscalar and scalar mesons above the chiral transition.

A more important effect is associated with the temperature dependence of the
anomaly term (\ref{lagsu3}).
If the anomaly term becomes considerably weaker 
as the temperature is increased,
it will reduce the thermal masses of the $\eta'$ rather than increase them.
The thermal mass of the $a_0$ mesons will also be reduced by this effect.
First estimates indeed show that this is the case,
so that these thermal fluctuations then ought to be included.
Whether the presence of these additional low-lying mass states
will change the order of the phase transition
is beyond the scope of the present investigation
but will be addressed in a forthcoming work.


\section{Dynamics}
\label{sec:dynamics}


In this section,
we consider a simplified dynamical scenario
that approximates the conditions
occurring in the central region of a relativistic heavy-ion collision.
For this purpose,
we adopt the approach given in Ref.\ \cite{Randrup:PRL},
in which a scaling expansion is emulated by means of a Rayleigh cooling,
and extend it to SU(3). 
We discuss the time evolution of the order parameters
as well as that of the pseudoscalar and scalar meson masses.


\subsection{Scaling expansion in SU(3)} 

As an initial state,
we consider thermalized matter at a specified temperature,
with only the scalar and pseudoscalar mesons included.
The matter is then assumed to experience a uniform scaling expansion
which causes the field fluctuations to decrease
according to the dimensionality $D$ of the expansion.
This brings the system out of equilibrium,
as the effective potential governing the order parameter
reverts to its vacuum form faster than the order parameter can follow.
For the standard SU(2) case,
a boost invariant expansion, corresponding to $D=1$,
was considered in Ref.\ \cite{Asakawa},
while the three-dimensional scenario was explored
in Refs.\ \cite{GM94,Lampert}.
If the expanding system is strictly uniform,
its properties depend only on the elapsed proper time $\tau$,
\ie\ the time experienced in a local comoving frame,
and the d'Alembert operator is then given by
\begin{equation}
\Box\ \longrightarrow\
{1\over\tau^D}{\partial\over\partial\tau}
\tau^D{\partial\over\partial\tau}\
=\ \frac{\partial^2}{\partial \tau^2} + \frac{D}{\tau}
\frac{\partial}{\partial \tau} \quad .
\end{equation}
The equations of motion of the two order parameters
are therefore readily obtained,
\begin{eqnarray}
\left(\frac{\partial^2}{\partial \tau^2} + \frac{D}{\tau}
\frac{\partial}{\partial \tau} + \mu_\sigma^2 \right) \sigma &=& 
f_\pi m_\pi^2 + c \left(\frac{4}{3\sqrt{2}} \delta\eta^2 
+ \frac{1}{\sqrt{2}} \delta K^2 + 2 \delta\sigma\delta\zeta \right) +
\cr && {}
\left(\sqrt{2} \lambda' \delta K^2 
- 8 \lambda (\delta\sigma\delta\zeta) \right)\zeta \label{eq:eomstau} \\
\left(\frac{\partial^2}{\partial \tau^2} + \frac{D}{\tau}
\frac{\partial}{\partial \tau} + \mu_\zeta^2 \right) \zeta &=& 
\sqrt{2} f_K m_K^2 - \frac{1}{\sqrt{2}} f_\pi m_\pi^2 
+ c \left(\delta\pi^2+\sigma^2+\delta\sigma^2-\frac{1}{3}\delta\eta^2 \right)
\cr && {}
+ \left(\sqrt{2} \lambda' \delta K^2 
- 8 \lambda \delta\sigma\delta\zeta \right) \sigma \label{eq:eomztau}
\quad ,
\end{eqnarray}
where the effective masses $\mu_\sigma$ and $\mu_\zeta$
are given in Eqs.\ (\ref{eq:musigma}) and (\ref{eq:muzeta}).

The equations of motion (\ref{eq:eomstau}-\ref{eq:eomztau}),
can be used to propagate the order parameter.
At each time step,
the effective masses are obtained by means of the expressions
given in the Appendix.
In order to approximately emulate the effect of the expansion,
we assume that the field fluctuations exhibit a simple power behavior,
\begin{eqnarray}
\delta\pi(\tau)^2 = 
\delta\pi(\tau_0)^2 \left(\frac{\tau_0}{\tau}\right)^D \; & , & \quad
\delta K(\tau)^2 = 
\delta K(\tau_0)^2 \left(\frac{\tau_0}{\tau}\right)^D \quad , \cr
\delta\eta(\tau)^2 = 
\delta\eta(\tau_0)^2 \left(\frac{\tau_0}{\tau}\right)^D \; & , & \quad
\delta\sigma(\tau)^2 = 
\delta\sigma(\tau_0)^2 \left(\frac{\tau_0}{\tau}\right)^D \quad ,
\end{eqnarray}
as they will eventually when the system has become so cold
that the individual quasi-particle modes are decoupled
($\tau_0$ denotes the initial time when the expansion is started).
Because of the mixing between $\delta\sigma$ and $\delta\zeta$,
and the interdependence of the various quantities,
a simple iteration procedure is carried out until convergence is achieved
at each time step.


\subsection{Evolution of the order parameter} 

\begin{figure}[htbp]
\begin{center}
\leavevmode
\epsfxsize=0.7\textheight
\epsfbox{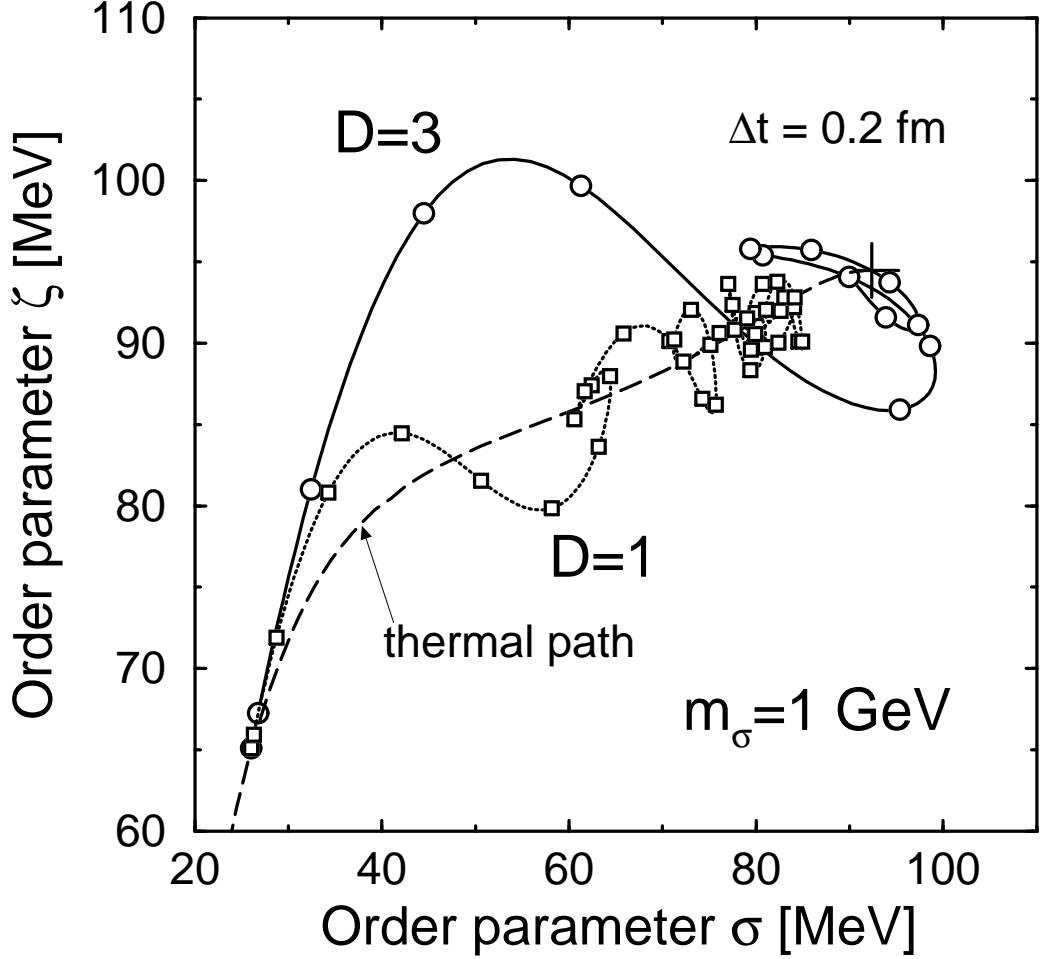}
\end{center}
\caption{The dynamical path of the order parameter $(\sigma,\zeta)$
as a result of a uniform pseudo-expansion in one and three dimensions,
starting from thermal equilibrium at $T=400~\MeV$.
Time steps with $\Delta\tau=0.2~\fm$ are indicated
along the trajectories with squares ($D$=1) and circles ($D$=3).
The thermal path is shown for comparison (long-dashed curve)
and the cross indicates the location of the vacuum,
$(\sigma_{\rm vac},\zeta_{\rm vac})$,
towards which all trajectories converge in time.}
\label{fig:dyn1000_order}
\end{figure}

In our first illustrations,
we employ a relatively high initial temperature, $T_0=400~\MeV$,
in order to achieve a high degree of chiral symmetry restoration
in the initial state.
Moreover, the initial time is chosen as $\tau_0=1~\fm$,
as is most commonly done,
and $\dot{\sigma}(\tau_0)=0$ and $\dot{\zeta}(\tau_0)=0$.
Figure \ref{fig:dyn1000_order}
shows the trajectories of the two order parameters for $m_\sigma=1~\GeV$
and in two different expansion scenarios,
a longitudinal ($D=1$) and isotropic ($D=3$).
The former represents a lower bound on the degree of expansion,
as it assumes that no transverse expansion occurs at all,
whereas the latter probably overestimates the expansion,
so the actual physical scenarios is expected
to be intermediate between these two extremes.

For the three-dimensional expansion,
the strange order parameter is changing first and overshooting its vacuum value.
The $\sigma$ order parameter remains quite unchanged during that time
and starts growing towards its vacuum value only at later times.
The long-dashed curve traces the movement of the equilibrium point
as the temperature is reduced;
this thermal path would be followed if the expansion were adiabatic
so the system had time to equilibrate throughout the expansion process.
However,
because the expansion is fast,
the dynamical trajectory oscillates around the thermal path.
For the fastest expansion,
only one oscillation occurs before the effective potential has
reestablished its vacuum form and the remaining oscillations
are then around the vacuum point.
It is interesting that the early oscillations of $\zeta$
are much stronger than those of $\sigma$,
whereas the situation is reversed
for the late oscillations around the vacuum point.

The one-dimensional expansion shows a distinctly different behavior.
As above,
the initial evolution is primarily in the direction
of the strange order parameter,
but there are now several oscillations around the adiabatic path
before the vacuum point is reached.
This is because the order parameter trajectory
oscillates around the instantaneous minimum in the effective potential
which now develops relatively slowly
from its initial value near the origin towards the vacuum point.
Thus there is more opportunity for the order parameter
to overshoot the vacuum point and rebound towards smaller values.

\subsection{Evolution of the thermal fluctuations} 

\begin{figure}[htbp]
\begin{center}
\leavevmode
\centerline{\epsfxsize=0.5\textwidth\epsfbox{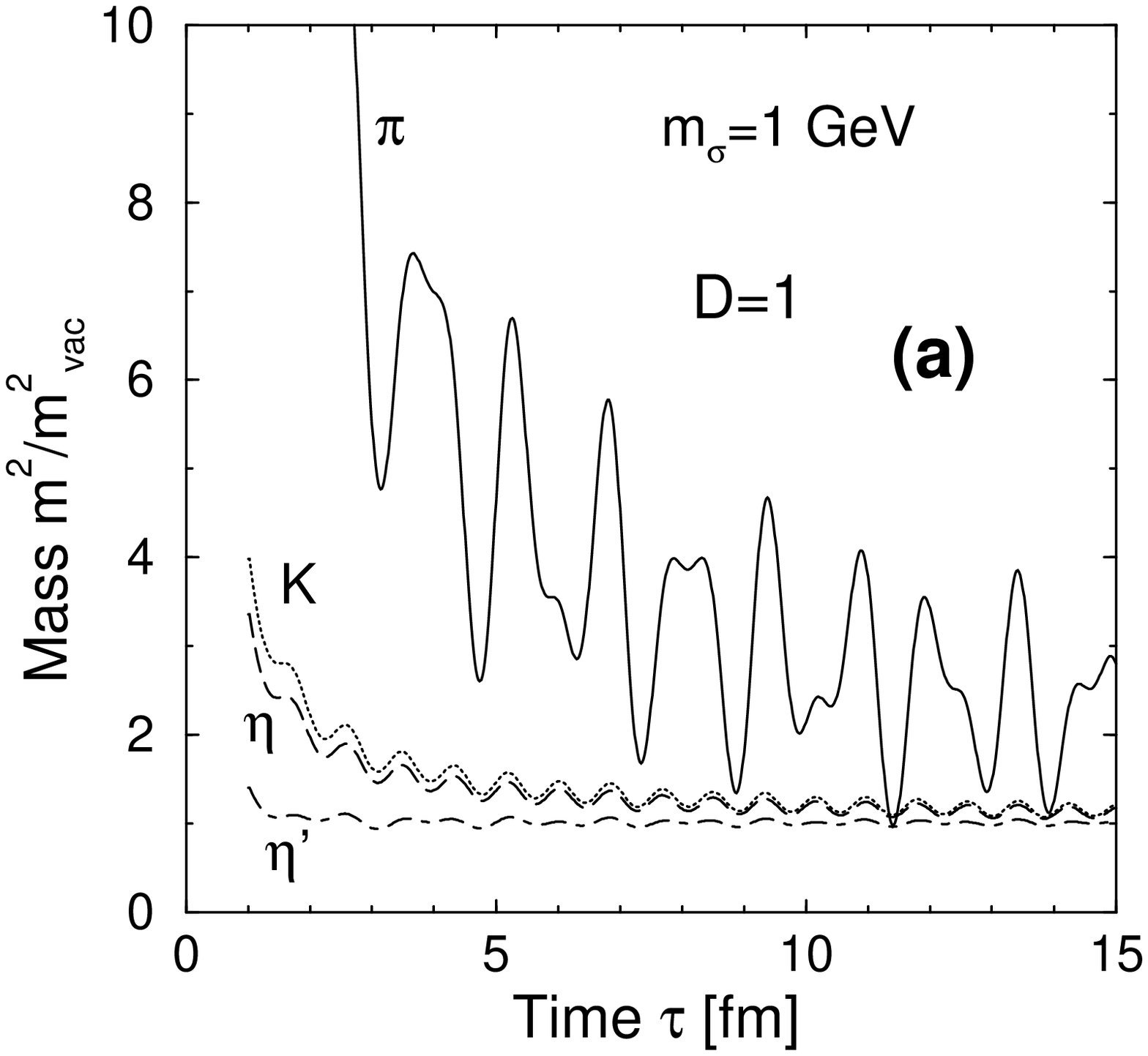}
\epsfxsize=0.5\textwidth\epsfbox{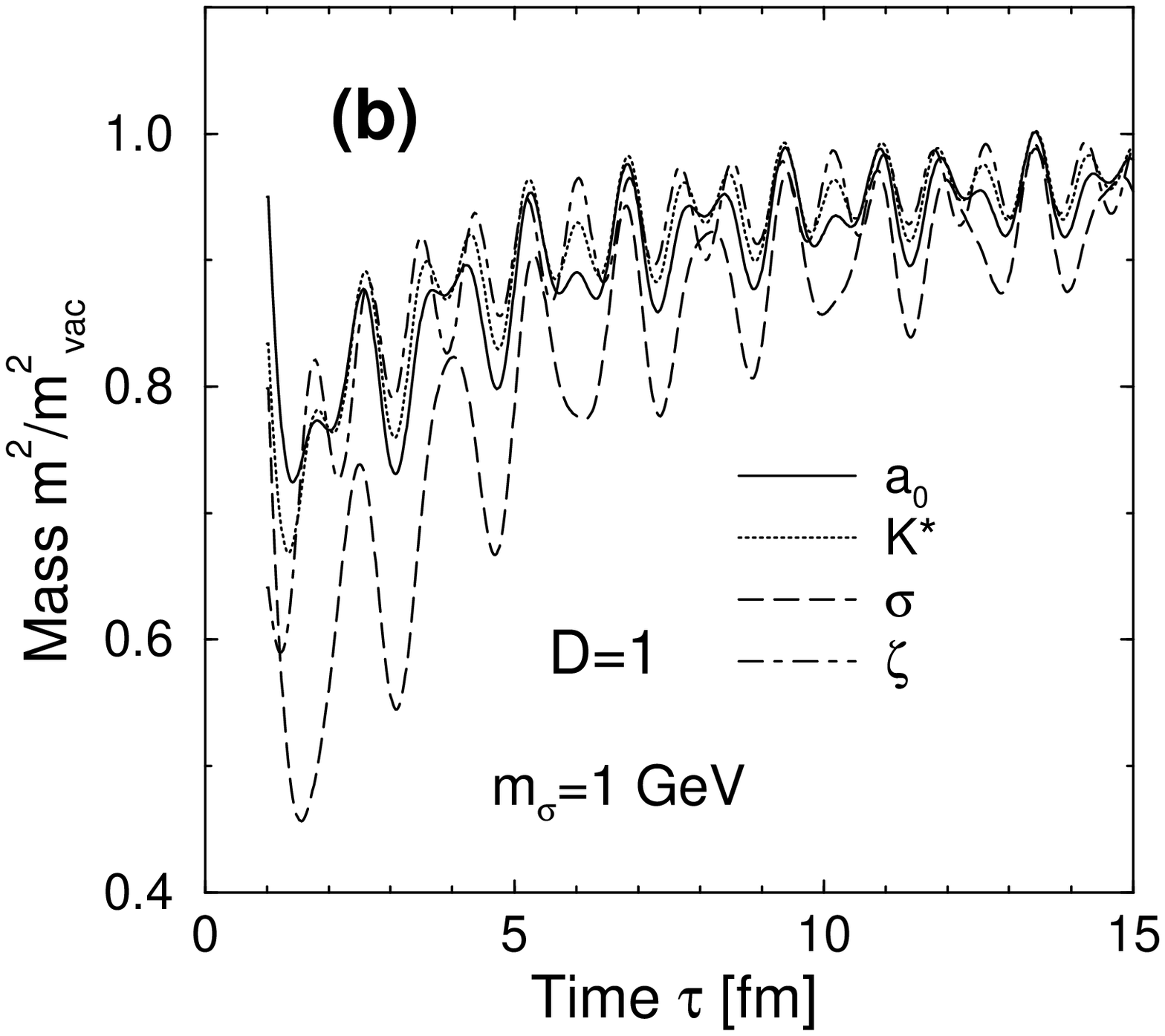}}
\end{center}
\caption{The time evolution of the (squares of the) effective masses 
(a) for the pseudoscalar mesons $\pi$, $K$, $\eta$, $\eta'$ 
and (b) for the scalar mesons $\sigma$, $\zeta$, $K^*$, $a_0$ 
in a pseudo-expansion with $D=1$.
The masses are given relative to their respective free values
and $m_\sigma=1~\GeV$ has been used.}
\label{fig:dyn_mass1d}
\end{figure}

The pattern in the evolution of the order parameters
will be reflected in the behavior of the pseudoscalar and scalar mesons masses.
Figure \ref{fig:dyn_mass1d} (left) shows the time dependence of
the pseudoscalar meson masses for the one-dimensional expansion. 
The masses are scaled to their respective vacuum values
so that they will tend to unity at large times.
The pseudoscalar meson masses remain well above zero during the evolution.
The initial value of the pion mass is much larger than its vacuum value
and it exhibits a slowly decreasing average behavior,
overlayed with large oscillations containing several different frequencies.
This rather complicated pattern is a result of
the coupled evolution of the two order parameters during the expansion.
The oscillations are nearly undamped over the entire time interval shown,
{\em i.e.}\ up to $\tau=15~\fm$.
The $K$ and $\eta$ masses also show oscillations
but they appear to have only one main frequency
and the amplitudes are much smaller than for the $\pi$ mass. 
The $\eta'$ mass does not change very much
and has only small fluctuations around its vacuum value.

The scalar masses experience a different time evolution,
as depicted in Fig.\ \ref{fig:dyn_mass1d} (right).
All scalar masses start below their vacuum value
and approach their free value from below,
contrary to the behavior of the pseudoscalar mesons. 
The oscillations of the scalar meson masses are sometimes in and sometimes out
of phase.
Especially the $\sigma$ and the $\zeta$ meson masses
are quite often in opposite phases.
The amplitudes of the oscillations are rather similar,
with that of the $\sigma$ meson mass being the largest,
and they are not sufficiently large to overshoot the respective vacuum value. 

Performing a Fourier transform of $m^2(\tau)$,
we find that there are two main frequencies involved,
one at $0.74~\fm^{-1}$ and the other at $1.2~\fm^{-1}$.
The oscillations of the $\sigma$ order parameter
is dominated by the lower frequency,
with a weaker component from the higher one,
whereas the strange order parameter has only the higher frequency. 
The pion has both frequencies with the lower one being stronger.
The $K$, $\eta$, and K$^*$ masses follow the frequency of the
strange order parameter,
while those of $\eta'$ and $a_0$ seem to have both frequencies
at about equal strength.

\begin{figure}[htbp]
\begin{center}
\leavevmode
\centerline{\epsfxsize=0.5\textwidth\epsfbox{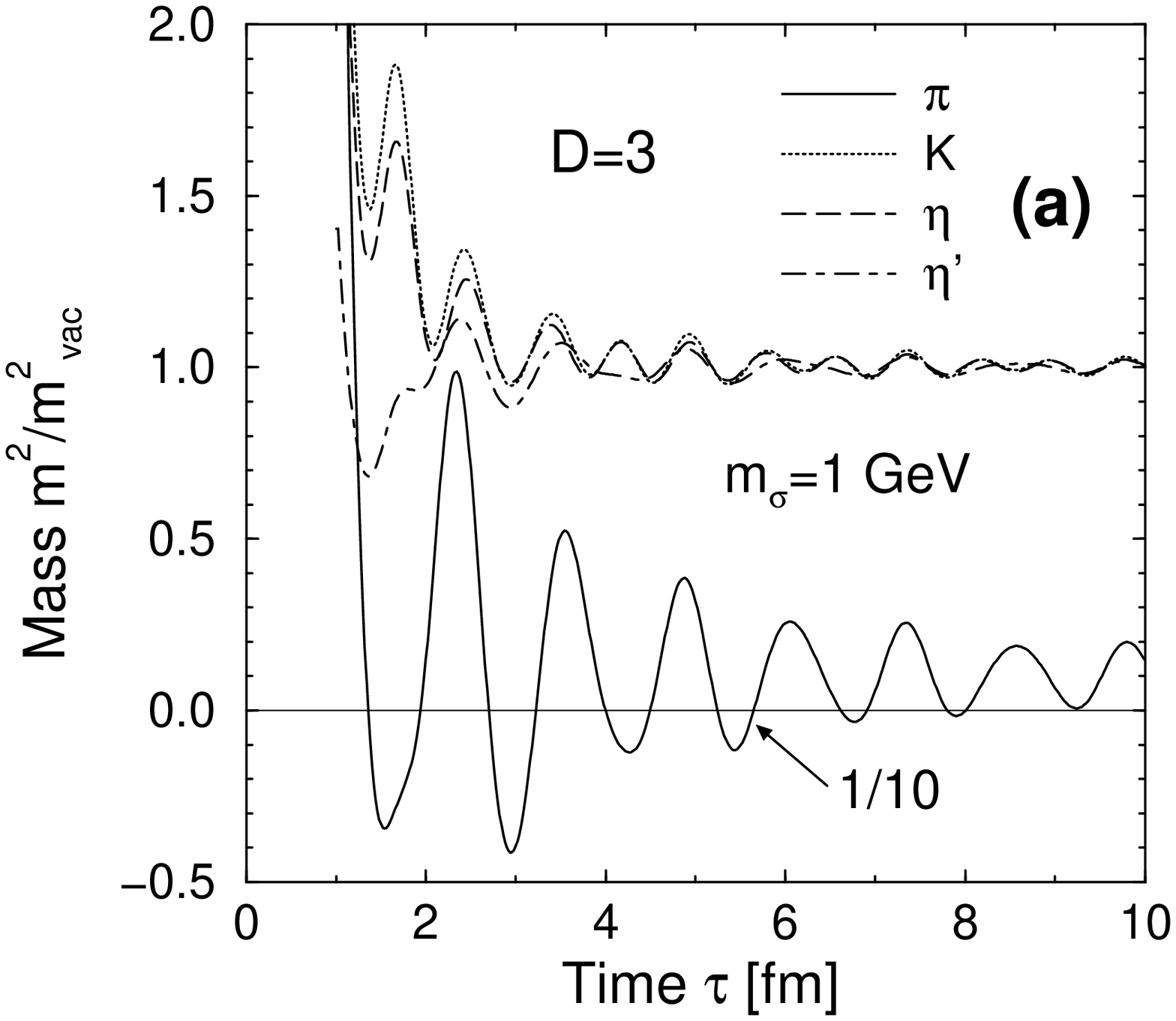}
\epsfxsize=0.5\textwidth\epsfbox{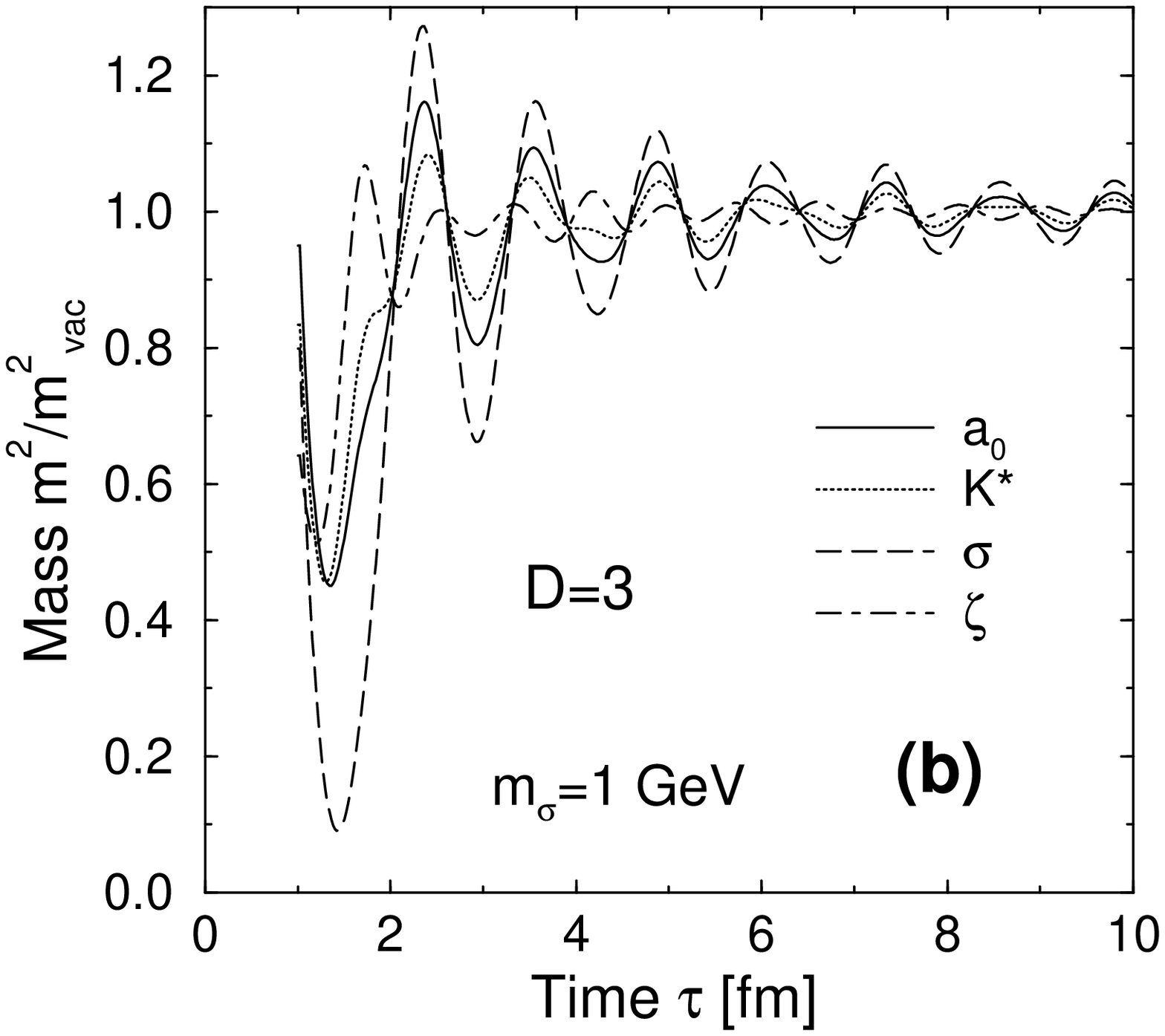}}
\end{center}
\caption{The time evolution of the (squares of the) effective masses 
(a) for the pseudoscalar mesons $\pi$, $K$, $\eta$, $\eta'$ 
and (b) for the scalar mesons $\sigma$, $\zeta$, $K^*$, $a_0$
in a pseudo-expansion with $D=3$.
The masses are given relative to their respective free values
and $m_\sigma=1~\GeV$ has been used.}
\label{fig:dyn_mass3d}
\end{figure}

Now we turn to the three-dimensional expansion.
Figure \ref{fig:dyn_mass3d} (left) shows the time dependence
of the pseudoscalar masses.
The oscillations of the pion mass are now much stronger
than for the other pseudoscalar mesons
(and the pion curve is therefore scaled down by a factor of ten).
The oscillations are also much smoother than in the one-dimensional case.
Most significantly,
the oscillations brings the pion mass into the region of instability
several times (a total of six in the specific calculation).
Remarkably,
the second minimum at $\tau\approx3~\fm$
is slightly deeper than the first minimum.
This is a result of the dynamical interplay between the two order parameters.
Indeed,
the order parameters are not really in phase at the time of the first minimum,
as the initial change of $\zeta$ is much larger than that of $\sigma$.
However,
at the time of the second minimum the two order parameters
are already oscillating around the vacuum point and are in phase,
so that the oscillations in the pion mass are enhanced.
At $\tau\approx3~\fm/c$ the masses of all the other pseudoscalar mesons
also exhibit their lowest minimum,
but the mass oscillations of these mesons
are weaker than in the one-dimensional expansion
(note the different mass scales in
Figs.\ \ref{fig:dyn_mass1d} and \ref{fig:dyn_mass3d}). 
The $K$ and $\eta$ masses are evolving very similarly,
but the $\eta'$ mass shows a different behavior:
it is not oscillating with one dominant frequency,
as was the case for the $\pi$, $K$, and $\eta$ masses.
For all pseudoscalar mesons,
the damping of the oscillations is stronger than in the longitudinal expansion,
as would be expected.

The scalar meson masses obtained for $D=3$
are plotted in Fig.~\ref{fig:dyn_mass3d} (right)
and their oscillations are seen to be stronger than for $D=1$. 
Again, the scalar meson masses start below their vacuum values
but then overshoot them later.
The $\sigma$ meson mass has a deep minimum at $\tau=1.4~\fm/c$
which comes remarkable close to zero.
Also the $a_0$ has strong oscillations in phase with the pion.
The frequencies of the $K$ and $\zeta$ masses
differ from those of the $\pi$ and $\eta$ masses. 

A Fourier decomposition of $m^2(\tau)$
reveals that there are again two main frequencies,
$0.78~c/\fm$ and $1.22~c/\fm$.
The $\pi$, $\sigma$, and $a_0$ have the lower frequency
and the K$^*$ and $\zeta$ have the higher one,
while the $\eta'$ has both frequencies with equal strength.
By contrast,
the $K$ and $\eta$ masses evolve with a broad distribution of frequencies
around $1.3-1.56~c/\fm$. 

Finally,
we note that since the displayed masses
are scaled relative to their free values,
the pion will always seem to have the strongest oscillations
which could be somewhat misleading.
In fact the pseudoscalar mass shifts are all of the same order of magnitude,
around 150--400 MeV for the first extremum in the evolution for $D=3$.
On an absolute scale,
the fluctuations of especially the $K$ and $\eta$ masses
are as large as those of the $\pi$ mass
which my possibly lead to observable effects
in relativistic heavy-ion collisions.

\subsection{Spectral enhancements} 

We finally utilize the calculated time-dependent effective $\pi$ and $K$ masses
to illustrate the possible associated enhancements in the corresponding
spectral distributions of the free mesons.
Towards this end,
we assume that the evolution of the field amplitude for a given meson mode
is governed by a Klein-Gordon field equation,
\begin{equation}
\ddot{\phi_\k}\ =\ \omega_k^2\ \phi_\k\quad ,\quad
\omega_k^2=m^2(t)+k^2\ .
\end{equation}
Here $\k$ denotes the momentum of the mode
and the time-dependent frequency is given by the indicated dispersion relation,
where $m^2(t)$ is the calculated time-dependent effective mass squared.
Invoking the method presented in Ref.~\cite{Randrup:X},
we can then calculate the enhancement coefficient for that particular mode,
$X_\k$.
This quantity (which is never less than unity)
expresses the degree to which the population of the mode
has been enhanced due to the time dependence experienced by the frequency.
Thus, if the initial occupation number is 
$\nu_\k(\tau_0)\equiv\langle\tau_0|a_\k^\dagger a_\k|\tau_0\rangle$,
then its final value is given in terms of $X_\k$ as
\begin{equation}
\nu_\k(\tau_f)\ \equiv\ \langle\tau_f|a_\k^\dagger a_\k|\tau_f\rangle\
=\ [\nu_\k(\tau_0)+\half]X_\k-\half,
\end{equation}
where the presence of the one half reflects the effect of
the quantum fluctuations of the field amplitude for that mode.

\begin{figure}[htbp]
\begin{center}
\leavevmode
\epsfxsize=0.5\textheight
\epsfbox{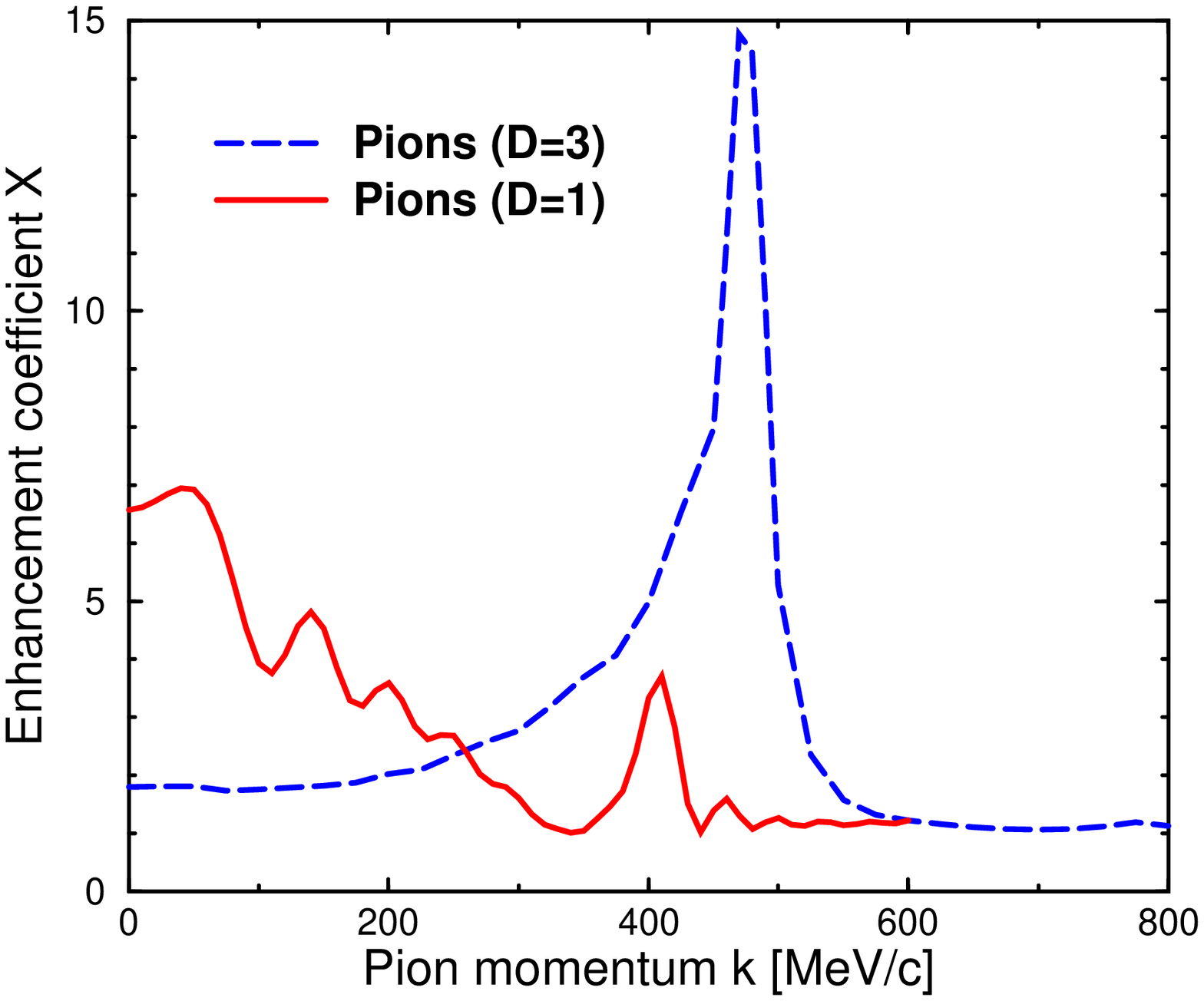}
\end{center}
\caption{The spectral enhancement coefficient $X_k$ for pions
as obtained on the basis of the calculated time-dependent effective masses
for the two pseudo-expansion scenarios considered,
in both cases starting from matter in thermal equilibrium at $T=400~\MeV$.}
\label{fig:Xpi}
\end{figure}

Figure \ref{fig:Xpi} shows the resulting pion enhancement coefficients
for the two pseudo-expansion scenarios considered.
For $D=1$ there is a significant enhancement of the softest modes
as well as a moderate peak near a pion momentum of 400~MeV.
The former feature is a direct reflection of
the relatively slow overall decrease of $m^2$ in the course of time,
while the peak near a momentum of $400~\MeV/c$
results from the quasi-periodic oscillations of the frequency.
This latter effect was discussed already in connection with the enhancement
of electromagnetic processes \cite{Boy:photons,KKRW}.
The results for $D=3$ exhibit a large and fairly sharp peak
at a slightly higher momentum,
with a more moderate and smooth enhancement of the soft modes.
The latter reflects the fact that $m^2(t)$ remains fairly constant on
the average, after the initial fast drop.
As a consequence of the more perfect quench achieved with $D=3$,
the oscillations of $m^2$ are more harmonic and, therefore,
the parametric amplification becomes stronger.
The resulting peak in the enhancement is located at a momentum of 480~MeV
which corresponds very closely to the expected resonance energy,
$\omega_1=\half m_\sigma=500~\MeV$.

\begin{figure}[htbp]
\begin{center}
\leavevmode
\epsfxsize=0.5\textheight
\epsfbox{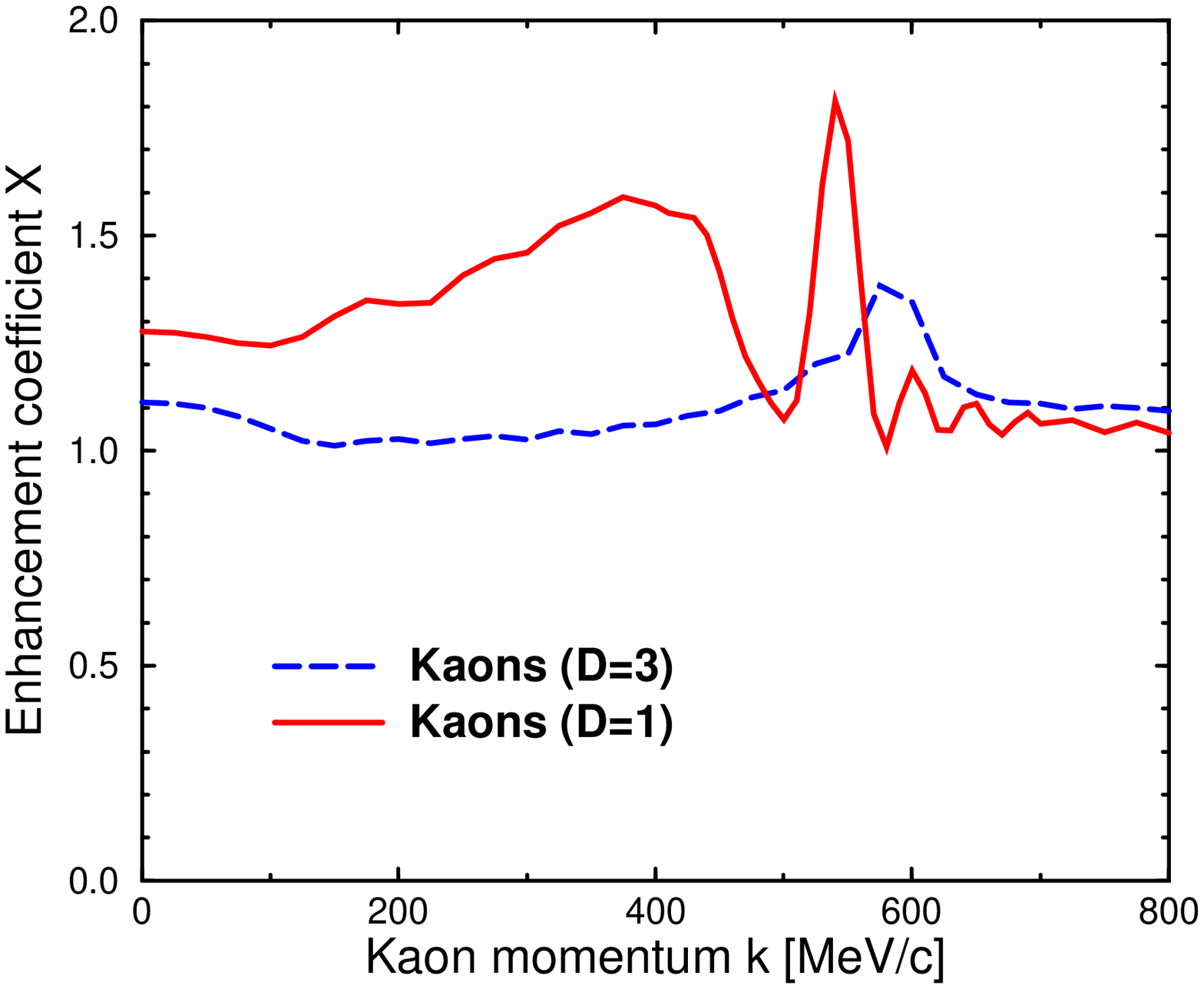}
\end{center}
\caption{The spectral enhancement coefficient $X_k$ for kaons,
as obtained on the basis of the calculated time-dependent effective masses
for the two pseudo-expansion scenarios considered,
in both cases starting from matter in thermal equilibrium at $T=400~\MeV$.}
\label{fig:XK}
\end{figure}

Figure \ref{fig:XK} shows the corresponding results for the kaons.
Since the relative modification of the effective kaon mass is
smaller (see Figs.\ \ref{fig:dyn_mass1d}-\ref{fig:dyn_mass3d}),
the associated enhancements are also much more modest.
Nevertheless,
it is possible to discern features similar to those of the pions.
In particular,
for $D=1$ there is relatively broad enhancement of the softer modes
followed by a small resonance peak and
for $D=3$ the most prominent feature is resonance
at a slightly larger momentum but,
perhaps surprisingly, it is quite small.

We finally wish to stress that these illustrations
are based on a fairly schematic picture
and they should not be taken as quantitative predictions.
Nevertheless, they do suggest that significant spectral
enhancements may be present in the SU(3) description
and they also give a general idea of their specific features.
Evidently, a quantitative prediction would require much more elaborate
calculations of the complicated collision dynamics.


\section{Conclusions}
\label{sec:conclusions}

The present investigation presents a first exploration
of how the inclusion of strangeness in the effective $\sigma$ model
affects the features associated with disoriented chiral condensates.
The resulting SU(3) model contains both the scalar and pseudoscalar
meson nonets and introduces the strange order parameter,
$\zeta=\langle\bar{s}s\rangle$,
thus making both the phase structure and the dynamics more complex.

The inclusion of strangeness is expected to have significant effects
once the temperature exceeds the mass of the strange quark
and thus it should not be ignored in DCC discussions.
Indeed,
the crossover towards chiral symmetry is being pushed to higher temperatures,
even for the non-strange order parameter $\sigma$,
though the equilibrium value of $\sigma$ is reduced
when the $\sigma$ mass is increased.
Consistent with this trend,
the thermal masses of the 18 mesons then also become more similar.

By considering suitable idealized expansion scenarios,
we have sought to ascertain the effect of the model extension
on the non-equilibrium dynamics following a high-energy nuclear collision.
The field fluctuations reflecting the presence of kaons
are important for the dynamics and can not be neglected.
Furthermore,
these calculations yield significant spectral enhancements 
not only for the pions (as was the case in the SU(2) description)
but also for the kaons (though to a much smaller degree).

We have pointed out that the neutral fraction $R_K$
of kaon emitted by a single DCC domain has a constant distribution,
$P_K(R_K)=1$
which is qualitatively as anomalous as the inverse square-root
distribution $P_\pi(R_\pi)=1/2\sqrt{R_\pi}$ 
governing the neutral fraction of DCC pions.
Relative to this latter distribution,
the kaon distribution $P_K(R_K)$ may be more amenable to measurement
by suitable event-by-event analysis
due to the fact that $K_S^0$ decays into charged pions.

Finally, if chiral symmetry is approximately restored, 
the $K/\pi$ and $\eta/\pi$ equilibrium ratios are enhanced by factors of 2-3
due to the mass shifts of the mesons.
If the expansion is sufficiently fast,
this effect may be observable,
even though the  signal will be reduced by decaying resonances.

The present investigation assumes that the coupling constants are temperature 
independent.
A possible temperature dependence of the anomaly term
will certainly change the meson mass spectrum at the higher temperatures,
as well as the evolution of the order parameters and the effective masses.
This will be studied in a forthcoming work.

Our present exploratory study suggest 
that various effects associated with the chiral phase transition in SU(3)
might be observable at RHIC
and we hope that our present findings 
will stimulate more elaborate investigations.


\acknowledgments

We are pleased to acknowledge helpful discussions with many
colleagues,
including Volker Koch, Ramona Vogt, Xin-Nian Wang, and Hans-Georg Ritter
at Lawrence Berkeley National Laboratory and
Panajotis Papazoglou, Detlef Zschiesche, Stefan Schramm, and Horst St\"ocker
at Frankfurt, Germany.
J.S.-B. acknowledges support by the Alexander-von-Humboldt Stiftung.
This work was supported by the Director,
Office of Energy Research,
Office of High Energy and Nuclear Physics,
Nuclear Physics Division of the U.S. Department of Energy
under Contract No.\ DE-AC03-76SF00098.

%
%

\section*{Appendix}

The effective masses for the pseudoscalar mesons
$\pi$, $K$, $\eta$, and $\sigma$ are given by
\begin{eqnarray}
m^2_\pi &=& -\mu^2 
+ 2\lambda\left(\frac{5}{3}\delta\pi^2 + \delta K^2 + \delta\eta^2 
+ \sigma^2 + \delta\sigma^2 \right) \cr &&{}
+ 4\lambda' \left(\frac{5}{3}\delta\pi^2 + \delta K^2 + \delta\eta^2 +
\sigma^2 + \delta\sigma^2 + \zeta^2 + \delta\zeta^2\right) \cr &&{}
-2 c \zeta \quad ,\cr\cr
m^2_K &=& -\mu^2 
+ 2\lambda\left(\delta\pi^2 + \frac{3}{2} \delta K^2 + \delta\eta^2 
+ \sigma^2 + \delta\sigma^2 + 2\zeta^2 + 2\delta\zeta^2 
- \sqrt{2}\left(\sigma\zeta+\delta\sigma\delta\zeta\right)\right) \cr &&{} 
+ 4\lambda' \left(\delta\pi^2 + \frac{3}{2}\delta K^2 + \delta\eta^2 + \sigma^2
+ \delta\sigma^2 + \zeta^2 + \delta\zeta^2\right) \cr &&{}
- \sqrt{2}c\sigma \quad ,\cr\cr
m^2_{P00} &=& -\mu^2 
+ 4\lambda\left(\delta\pi^2 + \delta K^2 + \delta\eta^2 
+ \frac{1}{3}\left(\sigma^2 +\delta\sigma^2 +
\zeta^2 + \delta\zeta^2\right)\right) \cr &&{}
+ 4\lambda' \left(\delta\pi^2 + \delta K^2 + \delta\eta^2 + \sigma^2 +
\delta\sigma^2 + \zeta^2 + \delta\zeta^2\right) \cr &&{}
+\frac{4}{3} c \left(\sqrt{2}\sigma+\zeta\right) \quad ,\cr\cr
m^2_{P88} &=& -\mu^2 
+ 2\lambda\left(\delta\pi^2 + \delta K^2 + 3\delta\eta^2 
+ \frac{1}{3}\left(\sigma^2 + \delta\sigma^2 +
4\zeta^2 + 4\delta\zeta^2\right)\right) \cr &&{}
+ 4\lambda' \left(\delta\pi^2 + \delta K^2 + 3\delta\eta^2 + \sigma^2 +
\delta\sigma^2 + \zeta^2 + \delta\zeta^2\right) \cr &&{}
-\frac{2}{3} c \left(2\sqrt{2}\sigma-\zeta\right) \quad ,\cr\cr
m^2_{P08} &=& 
\sqrt{2}\lambda\left(2\delta\pi^2 - \delta K^2 - 2\delta\eta^2 
+ \frac{2}{3}\left(\sigma^2 + \delta\sigma^2 -
2\zeta^2 - 2\delta\zeta^2\right)\right) \cr &&{}
-\frac{2}{3} c \left(\sigma-\sqrt{2}\zeta\right) \quad ,
\end{eqnarray}
where the thermal fluctuation of the pion field
is denoted by $\delta\pi^2$ and similarly for the other fields.
The mixing between the thermal fluctuations of $\eta$ and $\eta'$ are ignored
because of their small mixing angle.
For finite temperature,
one gets also a mixing between $\pi$ and $\eta_0$
due to the thermal kaon fluctuations.
This term vanishes in the mean-field approximation in the above expressions
but it would give a mixing term $m^2_{\pi\eta_0} = 2\sqrt{3}\lambda\delta K^2$. 
Since this term will be important only at high temperatures,
where the $\pi$ and $\eta'$ masses are closer in value
and the thermal kaon field fluctuations are sizable,
we ignore it here. 

Using the notation of the Particle Data Group \cite{PDG96},
we define the pseudoscalar mixing angle as
\begin{equation}
\pmatrix{\eta' \cr \eta} 
= \pmatrix{ \phantom{-}\cos\theta_p & \sin\theta_p \cr
 -\sin\theta_p & \cos\theta_p }
\pmatrix{ \eta_0 \cr \eta_8 } \quad .
\end{equation}
The physical masses of the $\eta$ and $\eta'$ are given by
the mixing formula \cite{Carr71c},
\begin{eqnarray}
m^2_\eta &=& m^2_{P88} \cos^2\theta_p + m^2_{P00}\sin^2\theta_p 
- m_{P08}^2 \sin 2\theta_p \quad , \\
m^2_{\eta'} &=& m^2_{P88} \sin^2\theta_p + m^2_{P00}\cos^2\theta_p 
+ m_{P08}^2 \sin 2\theta_p \quad ,
\end{eqnarray}
where the pseudoscalar mixing angle $\theta_p$ is fixed by
\begin{equation}
\tan 2\theta_p = \frac{2 m_{P08}^2}{m_{P00}^2-m_{P88}^2} \quad. 
\end{equation}
These mixing formulas relate in a unique way the eigenvalues of the 
mixed mass matrix to the physical unmixed masses $m_\eta$ and $m_{\eta'}$. 
Note that the physical $\eta$ state is then the one
closest to the octet state $\eta_8$, while the $\eta'$ is the state closest to
the singlet state $\eta_0$.
When the masses are defined in the above way,
then $\pi$ and $\eta'$ are degenerate
when the anomaly term vanishes ($c=0$).

The corresponding expressions for the scalar mesons are
\begin{eqnarray}
m^2_{a_0} &=& -\mu^2 
+ 2\lambda\left(\delta\pi^2 + \delta K^2 + \frac{1}{3}\delta\eta^2 
+ 3\sigma^2 + 3\delta\sigma^2 \right) \cr &&{}
+ 4\lambda' \left(\delta\pi^2 + \delta K^2 + \delta\eta^2 +
\sigma^2 + \delta\sigma^2 + \zeta^2 + \delta\zeta^2\right) \cr &&{}
+ 2c\zeta \quad ,\cr\cr
m^2_{K^*} &=& -\mu^2 
+ 2\lambda\left(\delta\pi^2 + \delta K^2 + \frac{7}{3}\delta\eta^2 
+ \sigma^2 + \delta\sigma^2 + 2\zeta^2 + 2\delta\zeta^2 
+ \sqrt{2}\left(\sigma\zeta+\delta\sigma\delta\zeta\right)\right) \cr &&{} 
+ 4\lambda' \left(\delta\pi^2 + \delta K^2 + \delta\eta^2 + \sigma^2
+ \delta\sigma^2 + \zeta^2 + \delta\zeta^2\right) \cr &&{}
+ \sqrt{2}c\sigma \quad ,\cr\cr
m^2_{S00} &=& -\mu^2 
+ 2\lambda\left(\delta\pi^2 + \delta K^2 + \frac{1}{3}\delta\eta^2 
+ 3\left(\sigma^2 +\delta\sigma^2\right) \right) \cr &&{}
+ 4\lambda' \left(\delta\pi^2 + \delta K^2 + \delta\eta^2 + 3\sigma^2 +
3\delta\sigma^2 + \zeta^2 + \delta\zeta^2\right) \cr &&{}
- 2c\zeta \quad ,\cr\cr
m^2_{S88} &=& -\mu^2 
+ 4\lambda\left(\delta K^2 + \frac{2}{3}\delta\eta^2 
+ 3\zeta^2 + 3\delta\zeta^2\right) \cr &&{}
+ 4\lambda' \left(\delta\pi^2 + \delta K^2 + \delta\eta^2 + \sigma^2 +
\delta\sigma^2 + 3\zeta^2 + 3\delta\zeta^2\right) \quad ,\cr\cr
m^2_{S08} &=& 
-\sqrt{2}\lambda\delta K^2 
+ 8\lambda' \left(\sigma\zeta+\delta\sigma\delta\zeta\right)
- 2c\sigma \quad .
\end{eqnarray}
The physical masses for the scalar mesons $\sigma$ and $\zeta$
are obtained in analogy with the $\eta$ and $\eta'$ described above,
where
\begin{equation}
\pmatrix{\sigma \cr \zeta} 
= \pmatrix{ \phantom{-}\cos\theta_s & \sin\theta_s \cr 
-\sin\theta_s & \cos\theta_s }
\pmatrix{ \sigma_0 \cr \sigma_8 }\ ,
\end{equation}
with $\theta_s$ being the scalar mixing angle.

%
%


\begin{thebibliography}{99}
\bibitem{Anselm88}
A.A. Anselm, Phys. Lett. B217, 169 (1989).

\bibitem{Anselm91}
A.A. Anselm and M.G. Ryskin, Phys. Lett. B266, 482 (1991).

\bibitem{Blaizot:PRD46}
J.-P. Blaizot and A. Krzywicki, Phys. Rev. D46, 246 (1992).

\bibitem{bj}
J.D. Bjorken, K.L. Kowalski, and C.C. Taylor,
7th Les Rencontres de Physique de la Valle d'Aoste:
Results and perspectives in particle Physics,
La Thuile, Italy, 7-13 March 1993;
Report SLAC-PUB-6109 (1993).

\bibitem{Rajagopal:NPB399}
K. Rajagopal and F. Wilczek, Nucl. Phys. B399, 395 (1993).

\bibitem{HW:ee} Z. Huang and X.-N. Wang, Phys. Lett. B383, 457 (1996).

\bibitem{Boy:photons}D.~Boyanovsky, H.J.~de~Vega,  R.~Holman, and S. Prem Kumar,
        Phys. Rev. D56, 5233 (1997).

\bibitem{KKRW}         Y.~Kluger, V. Koch, J. Randrup, and X.N. Wang,
                        Phys. Rev. C57, 280 (1998).


\bibitem{Rajagopal:QGP2}
K. Rajagopal, in {\sl Quark-Gluon Plasma 2},
ed.\ R. Hwa, World Scientific (1995).


\bibitem{BK:96}     J.P.~Blaizot and A.~Krzywicki,
                        Acta Phys. Polon. 27, 1687 (1996).

\bibitem{Bjorken}
J.D. Bjorken, Acta Phys. Polon. B28, 2773 (1997).

\bibitem{Rajagopal:NPB404}
K. Rajagopal and F. Wilczek, Nucl. Phys. B404, 577 (1993).

\bibitem{GGP93}
S. Gavin, A. Gocksch, and R.D. Pisarski, Phys. Rev. Lett. 72, 2143 (1994).

\bibitem{Huang:PRD49}
Z. Huang and X.-N. Wang, Phys. Rev. D49, 4335 (1994).

\bibitem{GM94}
S. Gavin and B. M\"uller, Phys. Lett. B329, 486 (1994).

\bibitem{Blaizot:PRD50}
J.-P. Blaizot and A. Krzywicki, Phys. Rev. D50, 442 (1994).

\bibitem{Kluger}
F. Cooper, Y. Kluger, E. Mottola, and J.P. Paz,
Phys. Rev. D51, 2377 (1995).

\bibitem{Boyanovsky:PRD51}
D.\ Boyanovsky and H.J.\ de Vega, Phys.\ Rev.\ D51, 734 (1995).

\bibitem{Asakawa}
M. Asakawa, Z. Huang, and X.-N. Wang, Phys. Rev. Lett. 74, 3126 (1995).

\bibitem{Csernai}
L.P. Csernai and I.N. Mishustin, Phys. Rev. Lett. 74, 5005 (1995).

\bibitem{MM}
S. Mr\'owczy\'nski and B. M\"uller, Phys. Lett. B363, 1 (1995).

\bibitem{Randrup:PRL}
J.~Randrup, Phys. Rev. Lett. 77, 1226 (1996).

\bibitem{Biro}
T.S. Biro, D. Moln\'ar, Z. Feng, and L.P. Csernai,
Phys. Rev. D55, 6900 (1997).

\bibitem{CGreiner}
T.S. Biro and C. Greiner, Phys. Rev. Lett. 79, 3138 (1997).

\bibitem{Camelia}
G. Amelino-Camelia, J.D. Bjorken, and S.E. Larsson
Phys. Rev. D56, 6942 (1997).

\bibitem{Abada}
A. Abada and M.C. Birse, Phys. Rev. D57, 292 (1998).

\bibitem{Gell60}
M. Gell-Mann and M. Levy, Nuovo Cimento 16, 705 (1960).

\bibitem{Levy67}
M. Levy, Nuovo Cimento 52, 23 (1967).

\bibitem{Gell64}
M. Gell-Mann and Y. Ne'eman, {\em The Eightfold Way} (Benjamin Inc., New York,
  1964).

\bibitem{Gasi69}
S. Gasiorowicz and D.A. Geffen, Rev. Mod. Phys. 41, 531 (1969).

\bibitem{Sche71a}
J. Schechter and Y. Ueda, Phys. Rev. D3, 168 (1971).

\bibitem{Carr71a}
P. Carruthers and R.W. Haymaker, Phys. Rev. Lett. 27, 455 (1971).

\bibitem{Carr71b}
P. Carruthers and R.W. Haymaker, Phys. Rev. D4,  1808  (1971).

\bibitem{Olsh71}
R. Olshansky, Phys. Rev. D4,  2440  (1971).

\bibitem{Sche71b}
J. Schechter and Y. Ueda, Phys. Rev. D3, 2874 (1971).

\bibitem{Chan74}
L.-H. Chan and R.W. Haymaker, Phys. Rev. D10, 4143 (1974).

\bibitem{Weinberg}
S. Weinberg, Phys. Rev. Lett. 17, 616 (1966).

\bibitem{Torn96}
N.A. T\"ornqvist and M. Roos, Phys. Rev. Lett. 76, 1575 (1996).

\bibitem{PDG96}
{Particle Data Group}, Phys. Rev. D54, 1 (1996).

\bibitem{Torn95}
N.A. T\"ornqvist, Z. Phys. C68, 647 (1995).

\bibitem{Torn97}
N.A. T\"ornqvist, 
7th International Conference on Hadron Spectroscopy, 
Upton, NY, 25-30 August 1997,
AIP Conference Proceedings, Vol.~432, p.~840 

\bibitem{Kaym85}
O. Kaymakcalan and J. Schechter, Phys. Rev. D31, 1109 (1985).

\bibitem{Sche80}
J. Schechter, Phys. Rev. D21, 3393 (1980).

\bibitem{Mish98}
I.N. Mishustin and O. Scavenius, hep-ph/9804338  (1998).

\bibitem{Papa98}
P. Papazoglou, S. Schramm, J. Schaffner-Bielich, H. St\"ocker, and W. Greiner,
Phys. Rev. C57, 2576 (1998).


\bibitem{Pisa84}
R.D. Pisarski and F. Wilczek, Phys. Rev. D29, 338 (1984).

\bibitem{Gavin94a}
S. Gavin, A. Gocksch, and R.D. Pisarski, Phys. Rev. D49, R3079 (1994).

\bibitem{Cole85}
S. Coleman, {\em Aspects of Symmetry} (Cambridge University Press, New York,
  1985).

\bibitem{Bura98}
L. Burakovsky and T. Goldman, Phys. Lett. B427, 361 (1998) 

\bibitem{Shuryak94}
E. Shuryak, Comments Nucl. Part. Phys. 21, 235 (1994)

\bibitem{Schaefer96}
T. Sch\"afer, Phys. Lett. B389, 445 (1996)

\bibitem{Wein90}
J. Weinstein and N. Isgur, Phys. Rev. D41, 2236 (1990).

\bibitem{Black98}
D. Black, A.H. Fariborz, F. Sannino, and J. Schechter, 
Phys. Rev. D58, 054012 (1998).

\bibitem{Kogut91}
J.B. Kogut, D.K. Sinclair, and K.C. Wang, Phys. Lett. B263, 101 (1991).

\bibitem{Randrup:PRD}
J. Randrup, Phys. Rev. D55, 1188 (1997).

\bibitem{Kapusta96}
J. Kapusta and D. Kharzeev, L. McLerran, Phys. Rev. D53, 5028 (1996).

\bibitem{Huang96}
Z. Huang and X.-N. Wang, Phys. Rev. D53, 5034 (1996).

\bibitem{Kharzeev98}
D. Kharzeev, R. D. Pisarski, M. H. G. Tytgat, Phys. Rev. Lett. 81, 512 (1998).

\bibitem{Meyer96}
H. Meyer-Ortmanns and B.-J. Sch\"afer, Phys. Rev. D53, 6586 (1996).

\bibitem{Lampert}
M.A. Lampert, J.F. Dawson, and F. Cooper, Phys. Rev. D54, 2213 (1996). 

\bibitem{Randrup:X}
J. Randrup, preprint LBNL-42616 (1998).

\bibitem{Carr71c}
P. Carruthers and R.W. Haymaker, Phys. Rev. D4, 1815 (1971).

\end{thebibliography}

%
%

%
%

\begin{table}
\caption{The meson masses in vacuum for the three parameter sets 
corresponding to $m_\sigma=400$, 600, 800, and 1000 MeV. The pseudoscalar
masses, the pseudoscalar mixing angle, and the masses of $a_0$ and $K^*$
are independent of that choice and are given in the first line.
The coupling constants and the masses of the $\sigma$ and $\zeta$ mesons
with their mixing angle $\theta_s$ are listed.
All masses are given in MeV.}
\begin{tabular}{ddddddd}
$m_\pi(138)$ & $m_K(496)$ & $m_\eta(547)$ & $m_{\eta'}(958)$ &
$\theta_p$ & $m_{a_0}(980)$ & $m_{K^*}(1430)$ \cr
\hline
138 & 496 & 539 & 963 & $-$5{\grads}.0 & 1029 & 1124 \cr
\hline\hline
$m_\sigma$(400-1200) & $m_\zeta(980)$ & $\theta_s$ & 
$\mu^2/|\mu|$ [MeV] & $\lambda$ & $\lambda'$ & $c$ [MeV] \cr
\hline
400 & 1199 & 20{\grads}.4 & $-$494.5 & 11.63 & -1.48 & 1698 \cr
600 & 1221 & 15{\grads}.4 & $-$342.3 & 11.63 & 0.344 & 1698 \cr
800 & 1278 & 4{\grads}.6 & 306.6 & 11.63 & 3.367 & 1698 \cr
1000 & 1546 & $-$21{\grads}.8 & 807.2 & 11.63 & 11.35 & 1698 
\end{tabular}
\label{tab:par}
\end{table}


                        \end{document}